\documentclass[graybox]{svmult}

\usepackage{type1cm}        
\usepackage{makeidx}         
\usepackage{graphicx}        
\usepackage{multicol}        
\usepackage[bottom]{footmisc}

\usepackage{newtxtext}       %
\usepackage[varvw]{newtxmath}       

\makeindex             

\begin{document}

\title*{The Einstein Probe Mission}
\author{Weimin Yuan\thanks{corresponding author}, Chen Zhang, Yong Chen and Zhixing Ling}
\institute{W. Yuan \at  National Astronomical Observatories, Chinese Academy of Sciences, Beijing, China, \email{wmy@nao.cas.cn}
\and C. Zhang \at National Astronomical Observatories, Chinese Academy of Sciences, Beijing, China, \email{chzhang@nao.cas.cn}
\and Y. Chen \at  Institute of High-Energy Physics, Chinese Academy of Sciences, Beijing, China, \email{ychen@ihep.ac.cn}
\and Z. Ling \at National Astronomical Observatories, Chinese Academy of Sciences, Beijing, China, \email{lingzhixing@nao.cas.cn}
}
%
%
\maketitle

\abstract*{}

\abstract{The Einstein Probe (EP) is a mission designed to monitor the sky in the soft X-ray band. It will perform systematic surveys and characterisation of high-energy transients and monitoring of variable objects at unprecedented sensitivity and monitoring cadences. 
It has a large instantaneous field-of-view (3,600 sq. deg.), that is realised via the lobster-eye micro-pore X-ray focusing optics. 
EP also carries a conventional X-ray focusing telescope with a larger effective area to perform follow-up observations and precise positioning of newly-discovered transients. 
Alerts for transient objects will be issued publicly and timely.
The scientific goals of EP are concerned with discovering faint, distant or rare types of high-energy transients and variable sources.
Within the confines of a modest-sized mission, EP will cover a wide range of scientific topics, from the nearby to high-redshift Universe.
The Einstein Probe is a mission of the Chinese Academy of Sciences, and also an international collaborative project. 
This paper presents the background, scientific objectives, and the mission design including the micro-pore optics and CMOS technologies adopted, the instruments and their expected performance, and the mission profile. 
The development status of the project is also presented.
}

\vspace{0.5in}


\section{Introduction}
\label{sec:intr}

\subsection{Background}

Since the early days of X-ray astronomy, it has been realised that the X-ray sky is dynamic, pervaded by variables, transients and even violent outbursts. 
The timescales of their varying radiations span a wide range from sub-seconds to years.
The studies of these spectacular sources have greatly shaped our understanding of the Universe. 
These sources, often involving extremely high energetics, also provide irreplaceable laboratories to explore the limits of physical laws and to study matter under extreme conditions.
The study of such objects and phenomena is an indispensable branch of time-domain astronomy.

There is a rich history in monitoring the X-ray sky.
Over the past decades, a number of space instruments with wide field-of-views (FoVs) or rapid scan capability---the so-called all-sky monitor (ASM),  had been built and flown (Holt and Priedhorsky 1987).
They have made the discoveries of various classes of highly variable sources and transients. 
Monitoring of the X-ray sky started with the Vela-5 series and Ariel-5 satellites. 
Though Vela-5B discovered the first gamma-ray burst (GRB), the ASM onboard Ariel-5, with pin-hole cameras of only small apertures and a large detector area by the standards of that time, was sensitive enough to monitor the sky on a fast cadence.
Ariel-5 discovered several novae, including the well-known low-mass X-ray binary A0620-00, and other transient X-ray sources.
More X-ray novae and transients were discovered with the successive ASMs, such as Ginga/ASM, RXTE-ASM, HETE-2, and Beppo-SAX/WFC.  
 
The Beppo-SAX mission first demonstrated the power of supplementing a quick-pointing and sensitive telescope to an ASM, by precisely locating the X-ray afterglow of GRBs with  
prompt follow-up X-ray observations.
This made it possible for ground-based optical follow-up observations, leading to the settlement of the long-debated cosmic origin of GRBs (Costa et al. 1997). 
The Neil-Gehrels Swift Observatory, with the remarkable success of its XRT, highlighted such a strategy.

Narrow-field telescopes also contributed to the discovery of transients and flaring events in their sky surveys,  serendipitous observations, and even spacecraft slews from one pointing to another.
The ROSAT all-sky survey (RASS), though with poor sampling cadences, caught the high state of some transient sources, including tidal disruption events (TDEs)  (e.g. Komossa \& Bade 1999) predicted theoretically.  
In this way, some transients and highly variable sources were also discovered with X-ray telescopes such as XMM-Newton, Chandra, Swift/XRT, and eROSITA, despite their even smaller FoVs compared to that of ROSAT.  

In recent years, Integral (Winkler et al.\ 2003), Swift (Gehrels et~al.\ 2004), and MAXI (Matsuoka et~al.\ 2009) have greatly expanded our horizon in monitoring the X-ray sky and advanced our knowledge about the dynamic Universe.
New transients continue to be discovered, including new X-ray binaries,  
relativistic TDEs, supernova shock breakouts, high-redshift GRBs, and more (see e.g. Gehrels \& Cannizzo 2015 for a review).
These sources are thought to be of merely the tip of iceberg, and more events are needed to characterise their properties and to understand their nature.
However, the bulk of these extragalactic transients are distant and faint.
To detect them in large numbers requires better sensitivity than those currently available, by at least one order of magnitude. 

Most of the high-energy monitoring instruments built so far were/are operating in the medium-to-hard X-ray (above several keV) and gamma-ray bands. 
The soft X-ray sky, from several keV down to a few hundred eV, has been lacking extensive monitoring with sufficiently high sensitivity and cadence. 
A soft X-ray ASM will open up a new window of sky monitoring. 
Moreover, almost all the X-ray ASMs were flown as supplementary payloads or for specialised objects (e.g. HETE-2 and Swift for GRBs).
There has been a long need for an observatory-type mission dedicated to surveys of X-ray transients in general, with fast follow-up and tele-communication capabilities.  

It is remarkable that the current and future time-domain observatories operate across the entire electromagnetic (EM) wave spectrum, including Pan-STARRS, LSST, ZTF, CHIME, SKA, LOFAR and more. 
Some intriguing transient events have also been discovered, such as fast radio bursts (FRB). 
X-ray observations played a key role in identifying the magnetar nature of FRB in at least one case (e.g. Li et al. 2021). 
Many new discoveries are anticipated when the sky is observed in time-domain simultaneously across a wide range of the EM spectrum.  

Last but not the least, astronomy has come of the long-awaited multi-messenger era.
Observatories such as LIGO, Virgo, Ice-cube, LHAASO, and CTA are extending our sight beyond only photons into gravitational waves (GW), neutrinos, and cosmic rays. 
The most remarkable is the GW event GW\,170817 resulting from the merge of two neutron stars detected with LIGO/VIRGO, together with its multi-wavelength counterparts detected by many telescopes (Abbott et al. 2017).
It opened up a completely new frontier, but its advance depends on detections of more of such events. 
Look into the future, with the advent of multi-wavelength and multi-messenger time-domain observatories, X-ray all-sky monitoring will enjoy a revival. 

 \subsection{Scientific motivations}

A tidal disruption event occurs when a star is getting too close to a massive black hole and gets tidally disrupted and accreted, producing a flare of radiation peaking in the UV and soft X-rays (Rees 1988).  
TDE provides a unique probe of quiescent massive black holes in normal galaxies, the accretion process throughout a complete cycle of accretion on a practically observable timescale, and the formation of relativistic jets. 
Recent observations have hinted at diverse types and properties of TDEs (e.g. Komossa 2017); however, further advance is hampered by the relatively small sample size, particularly in X-ray.
The newly launched eROSITA (Predehl et al. 2021) will detect many TDEs\footnote{A small, preliminary sample of eROSITA TDEs has been published (e.g. Sazonov et al. 2021).}, but the majority are expected to be faint and may not be found at the early or peak phase.  
Observations in the early and peak phase are essential for understanding some of the key  processes in TDE, and to understand the dichotomy of the X-ray and optical subclasses.
  
The early Universe at redshifts between 20 to 10 is thought to be the epoch when the first generation of luminous stars began to form and re-ionise the dark Universe. 
The only way to observe these youngest objects is to detect their explosive deaths, as GRB that is predicted.
They also signify the formation of the first black holes as remnants of stellar evolution.
High-redshift GRBs are therefore cosmic beacons to probe the first stars and black holes and the dark Universe.
They are faint, however, and require X-ray sensitivity $<1$~mCrab for the burst durations of the order of $10^4-10^5$ seconds.
Currently, only seven GRBs were detected with Swift at redshifts between 6.2 and 9.4  (e.g. Gehrels \& Cannizzo 2015).

Shock breakout from supernovae is the first X-ray and UV radiation produced as the outward-propagating shocks generated by core-collapse break out of the stellar surface (e.g. Falk 1978). 
They have timescales as short as the order of from 10s to $10^3$~seconds and moderate brightness of bursts.
Hence they are elusive, with only a few cases known so far (Soderberg et al. 2008).
Their observations can be used to constrain some of the key physical properties of core-collapse supernovae and their progenitor stars.

There also exist other various types of transient sources that are of no less scientific significance. 
Many await detection in larger samples and characterisation with data of better quality. 
Some examples include stellar flares, X-ray flashes, GRBs of various flavours and their precursors, magnetars, classical novae, supergiant fast X-ray transients, and outbursts of AGNs and blazars.
This diverse family is recently joined by a few quasi-periodic eruption sources (QPE, e.g. Arcodia et al. 2021), whose origins remain unknown.   

The detection of EM counterparts provides independent evidence for the physical processes involved in GW events.
It also enables the identification of the associated galaxies and study of the astrophysical properties of GW sources (Metzger 2017), leading to the redshift measurement.
The measured redshift, combining with the distance parameter inferred from the GW data, can be used as a cosmological probe to measure the geometry of the Universe. 
With the continuing upgrade, the GW observatories will detect many more events, calling for synergetic observations to detect their potential EM counterparts.
In X-ray, they may be detected as short GRB afterglows (as in GW\,170817) and other possible transient radiation predicted by models (e.g. Zhang 2013, Sun et al. 2019). 
Due to the poor localization capability of GW detectors (10--100 square degrees), rapid detection of the EM counterpart of GW events in the first place requires the instrument to have a large FoV.

\subsection{The Einstein Probe mission}

The Einstein Probe (EP) is a space mission dedicated to time-domain astronomy to monitor and characterise the soft X-ray transient sky (Yuan et al. 2015, 2016, 2018).
EP has a wide FoV of $\sim$3,600 sq.\,deg., combined with a high sampling cadence of several revisits per day for the night sky. 
EP also carries a conventional X-ray focusing telescope with a larger effective area to perform follow-up observations and precise positioning of newly-discovered transients. 
Transient alerts will be issued publicly and rapidly, in order to trigger multi-wavelength follow-up observations from the worldwide astronomical community. 
 
The Einstein Probe is a mission of the Chinese Academy of Sciences (CAS). 
It was first proposed in 2013 and formally adopted in December 2017.
The project is currently in the flight model phase (phase D). 
EP is planned for launch by the end of 2023. 
It has a lifetime of three years and five years as goal.
The Einstein Probe is an international collaborative project, participated formally and contributed by the European Space Agency (ESA, as Mission of Opportunity) and the Max-Planck-Institute for extraterrestrial Physics (MPE) in Germany.
A collaboration with the Centre National d'Etudes Spatiales (CNES) in France is also envisaged. 

This Chapter provides a review of the Einstein Probe mission. 
Section~\ref{sec:sci} summarises the science objectives and science requirements. 
In Section~\ref{sec:tech}, the new technologies that EP adopts, namely the lobster-eye optics and CMOS, are briefly introduced.
The design and predicted performance of the instruments are presented in Section~\ref{sec:inst}.
The satellite and mission profile are introduced in Section~\ref{sec:miss}.

\section{Science objectives}
\label{sec:sci}

Einstein Probe will carry out sensitive and systematic wide-field sky monitoring surveys in the soft X-ray band, which has not been fully explored in the time domain.
This is complemented by prompt and deep X-ray follow-up capability.
The primary science objectives are: 

\begin{enumerate}
\item{Discover and characterise cosmic X-ray transients, and monitor the variability of known sources, to reveal their properties and gain insight into their nature and underlying physics.}
\item{Discover and characterise X-ray outbursts from normally quiescent black holes, for better understanding of the demography of black holes and their formation and evolution, as well as accretion physics.}
\item{Search for X-ray sources associated with gravitational-wave events and precisely locate them.}
\end{enumerate}

Science objective~1 is what Einstein Probe is designed to achieve in the first place.
The increased sensitivity will enable unprecedented census of faint or distant high-energy transients beyond the horizon of previous and contemporary X-ray monitors. 
The combination of high sensitivity, large FoV (thus large grasp), and high cadence of EP will allow a large variety and number of X-ray transients to be detected, as well as new classes to be discovered. 
These will also allow transients to be found early in their outbursts, a time which has been generally inaccessible to the previous and existing missions.
The soft X-ray window implies that the bulk of the targets and processes being monitored will probably change from the non-thermal, highly relativistic beaming regime (e.g. GRBs, blazars) to the thermal, low-/non-beaming regime (e.g. accretion, shock), where new types of transients are expected to emerge. 

EP has a good prospect to detect GRBs at high redshifts, though the identification and redshift measurement of their host galaxies present great challenges.
EP is expected to detect more shock breakout events, yielding important clues to the properties of the progenitor stars. 
Other transient sources to study include X-ray flaring stars, X-ray flashes, various types of GRBs and GRB precursors, magnetars, classical novae, SFXT, and outbursts of active galactic nuclei and blazars.

Two other primary science objectives were identified---TDEs and the X-ray counterparts of gravitational-wave events. 
With the well suited soft X-ray bandpass,
EP is expected to detect and characterise TEDs in a large number and catch possibly some of them at the peak or even early phase of the flares. 
This will advance our understanding of the demography, formation, and evolution of massive black holes, as well as the physics of accretion and jet formation.
EP has the potential of detecting possible X-ray sources resulting from neutron star merger events, as short GRB afterglows and other possible transient flares. 
However, the objective is highly challenging as little is known about photonic emission in the soft X-ray band in the merging process of compact stars producing gravitational waves.

EP will monitor the variability of many X-ray sources all over the sky through its time-domain survey on a wide range of timescales from hours to years. 
Relatively faint X-ray sources, such as AGNs and ULXs in nearby galaxies, will be monitored routinely in larger numbers than what was accessible by the previous and current missions. 

EP will provide the X-ray counterparts or upper limits for transient sources discovered in other wavebands, as well as X-ray timing and spectral data for multi-wavelength variables, by operative synergies with other facilities over multi-wavebands.
A detection of the X-ray source of other transient multi-messenger events (neutrinos, high-energy gamma-rays and cosmic rays) would be of particular interest.
Some of such events may be associated with outbursts of blazars and supernova explosions.

\section{New technologies employed}
\label{sec:tech}

In  X-ray, the above driving science invokes  the next generation of instruments 
with higher sensitivity and angular resolution (several arcminutes or better) than those currently available.
The current X-ray monitors, which are all of non-focusing optics, have reached almost the limits imposed by the resources of space engineering (e.g. several hundred kilograms in weight) and the financial budget. 
The most efficient way to increase the signal to noise ratio, and thus the detection sensitivity, of an observation is to do imaging by focusing the X-ray light from a source. 
This can now be achieved by novel X-ray focusing technique---the lobster-eye micro-pore optics (MPO).
With X-ray focusing imaging, such optics can result in an enormously enhanced gain in signal to noise, and thus high sensitivity, while a wide FoV can be maintained. 

X-rays can only be reflected and focused under grazing incidence, with the  incidence angles smaller than a few degrees. 
The commonly used optics is the Wolter-I type, for which X-rays are reflected twice by a tubular parabolic surface followed by a hyperbolic surface.
Such optics, being rotationally symmetric, has naturally an optical major axis. 
When the off-axis angle of a light ray to the major axis gets larger, the imaging quality worsens quickly.
Consequently, Wolter-I telescopes have a small FoV, typically less than one degree.

The lobster-eye micro-pore optics mimics the imaging principle of the eyes of lobsters, which can make images by reflection instead of refraction. 
Although this idea has been put forward over 40 years (Angel 1979), its applications to X-ray astronomy proved to be a long and hard struggle.
Over the past decade or so, significant progresses have been made in the development of the MPO plates that suffice applications to wide-field X-ray imaging observations in space.

As for all wide-field instruments with moderate/good spatial resolution, large-format detectors, usually position-sensitive or pixelised, are required.
Position-sensitive gas proportional counters, which can be made in a large size, have been widely used. 
As working-horses in the history of X-ray astronomy, they are well matured technologically and easy to operate in orbit.
One good example is the Gas Slit Camera of MAXI (Mihara et al. 2011). 
The disadvantages of gas detectors are also obvious, however, mainly of fragility and gas leakage, and relatively low energy resolution.
The former presents a challenge and risk to the engineering, and the operation in space, of detectors sensitive in the soft X-ray band, since a very thin entrance window\footnote{The transmission efficiency for polymide with a thickness of 1~$\mu$m is only $\sim20\%$ at 0.5~keV.} has to be used. 

Alternative candidate detectors include micro-channel plates (MCP) and silicon-based CCD and CMOS sensors. 
On the basis of the overall evaluation considering the mission requirements, engineering and budget constraints, as well as the results of extensive laboratory tests, the scientific CMOS is used as the focal detectors for WXT. 
To our knowledge, this will be perhaps the first time that the CMOS is used as X-ray astronomical detectors in space. 
The characteristics and performance of the CMOS are briefly discussed  in Section~\ref{sec:cmos}.

\subsection{Lobster-eye micro-pore optics}
\label{sec:mpo}

\begin{figure}
 \sidecaption[b]
    \includegraphics[width=0.6\textwidth]{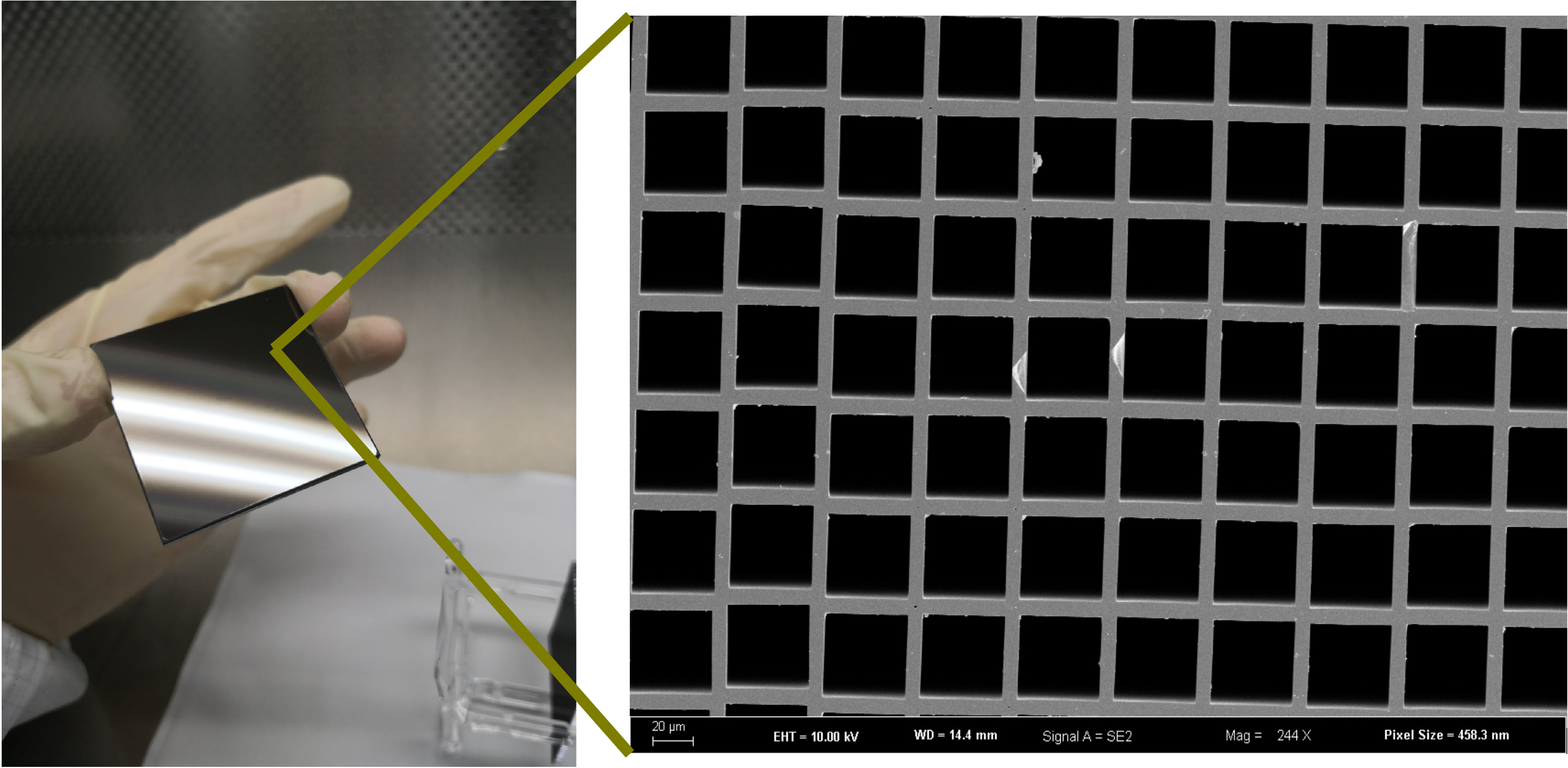}
    \caption{A MPO optic, with its array of pores revealed under microscope.}
    \label{fig:secmpo_photo}
\end{figure}

\begin{figure}
  \sidecaption[b]
    \includegraphics[width=0.5\textwidth]{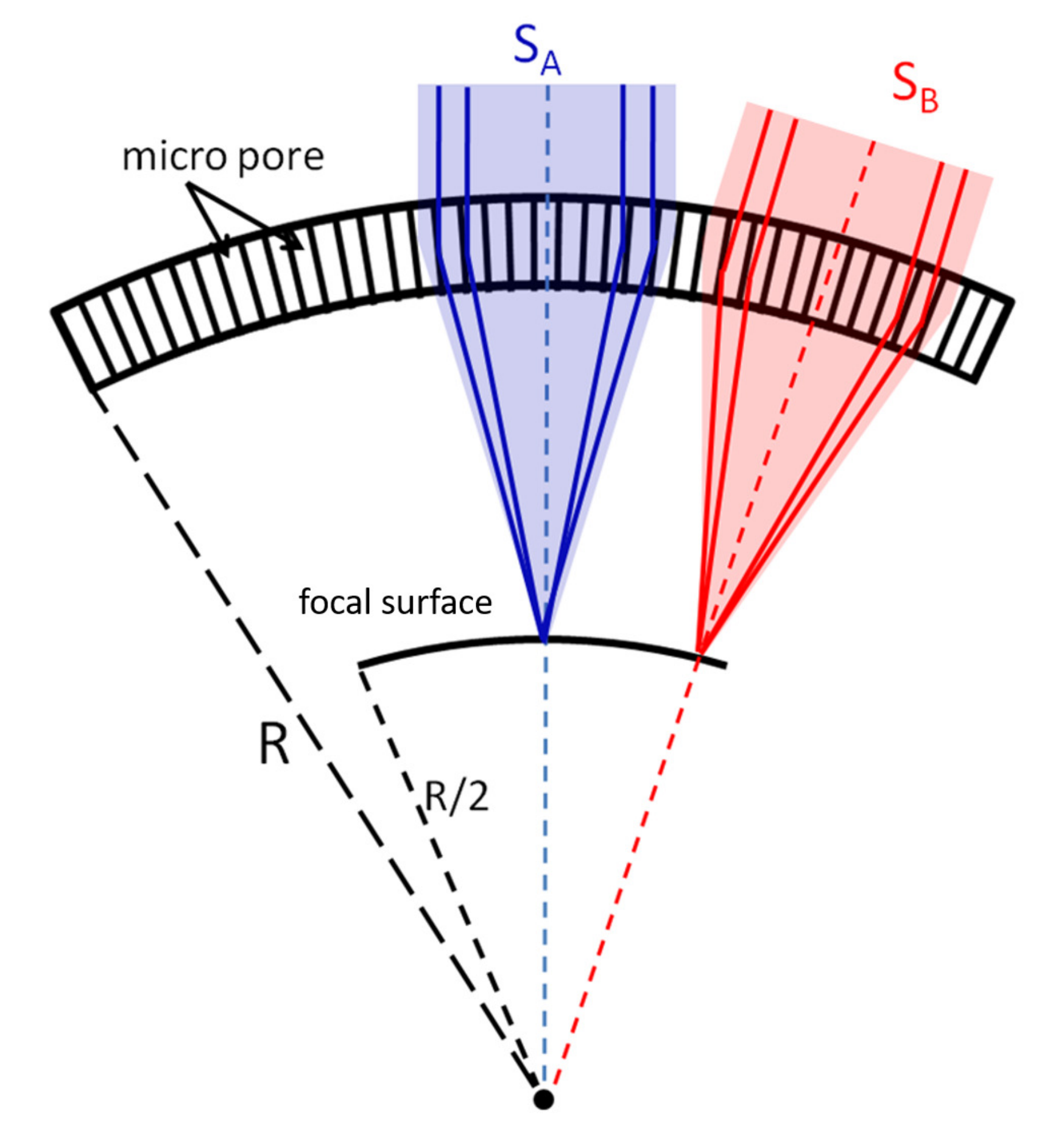}
    \caption{Imaging principle of lobster-eye micro-pore optics using a spherically slumped optic. X-ray photons from sources (A and B) in different directions are focused onto the two corresponding locations on the focal surface, by grazing-incidence reflections on the walls of the square pores.}
    \label{fig:mpo_principle}
\end{figure}

\begin{figure}
   \sidecaption[]
    \includegraphics[width=0.6\textwidth]{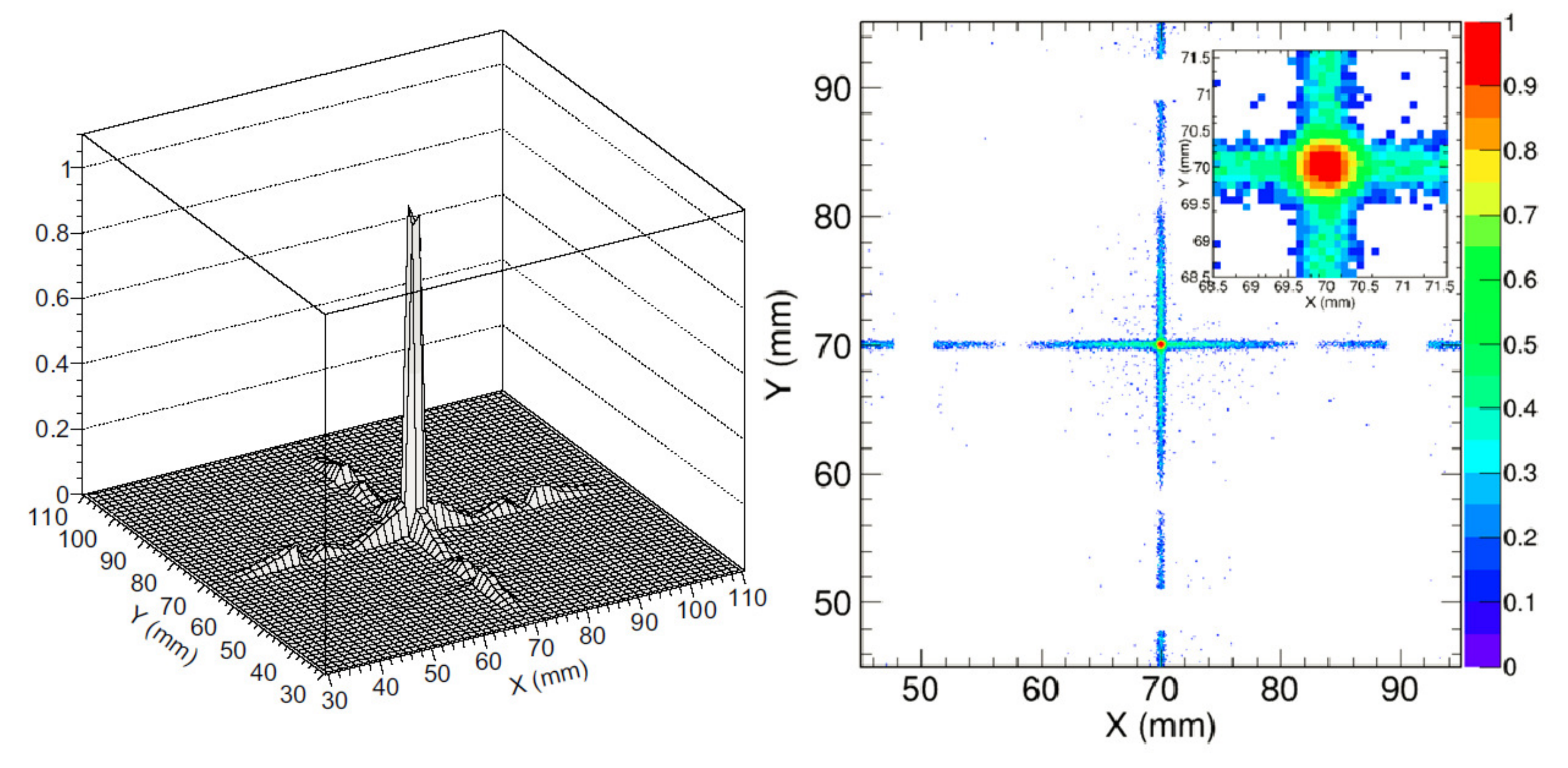}
    \caption{PSF of lobster-eye micro-pore optics for 1~keV photons derived from Monte-Carlo simulations (figure from Zhao et al. 2017).}
    \label{fig:mpo_psf}
\end{figure}

A lobster-eye MPO optic 
is made of a thin plate with millions of square micro-pores perpendicular to the surface, slumped into a spherical shape (Fig.~\ref{fig:secmpo_photo}).
The axes of all pores point radially to the curvature centre of the sphere, and the adjacent walls of pores are orthogonal to each other. 
As shown in Fig.~\ref{fig:mpo_principle}, incoming X-rays at a grazing-incidence angle are reflected off the walls of the pores, and are brought onto a focal sphere with a radius of half the curvature of the optic.  
It produces true imaging with a characteristic cruciform point-spread function (PSF), as shown in Fig.~\ref{fig:mpo_psf}.
The PSF consists of a central peak spot containing X-ray photons undergoing two reflections from adjacent walls of the pores and two cross-arms produced by photons with single reflection (and with odd-number reflections).
For an idealised lobster-eye optic, a same large number of pores take part in the reflections for any direction of incoming light.
This means that such an optic has no preferential major optical axis, and can have an almost un-vignetted, large FoV, up to the entire solid angle of $4\pi$.
Thus, in principle, the lobster-eye MPO provides a unique technique for wide-field X-ray focusing imaging.

Angel (1979) first proposed that the imaging principle of lobster's eye can be applied to build X-ray ASM.
Since then, this technique has been studied both theoretically and experimentally by a number of groups for many years (e.g. Wilkins et al. 1989, Chapman et al. 1991, Fraser et al. 1992, Kaaret et al. 1992, Willingale et al. 2016). 
NAOC has joined this effort since 2010 (Zhao et al. 2014). 
A few X-ray monitoring mission concepts were proposed as satellites  (e.g. Priedhorsky et al. 1996, Yuan et al. 2015) or piggyback payloads 
(e.g. Fraser et al. 2002, O'Brien 2021).  
The first real lobster-eye soft X-ray imager, STORM, was flown as a piggyback experiment on the sounding rocket mission DXL (Collier 2015).
The Mercury Imaging X-ray Spectrometer (MIXS, Fraser et al. 2010), built from MPO plates and aboard the already-launched BepiColombo to Mercury, will start observation in a few years. 
The narrow-field X-ray telescope MXT onboard the GRB mission SVOM is also built with MPO optics (G\"otz et al. 2014). 
For the currently commercially available MPO plates, a large focusing gain of $\sim2,000$ and moderate resolution of several arcminutes can be achieved.

\subsection{CMOS detectors}
\label{sec:cmos}

Complementary Metal-Oxide-Semiconductor (CMOS) sensors have been widely used  as industrial and commercial electronics in the past decade. 
Thanks to the rapid development of semiconductor manufacturing, the performance of CMOS sensor has been improved significantly and scientific CMOS sensors have also been developed\footnote{There have been long efforts in the development of CMOS for astronomical applications by several groups, including that described here and the hybrid CMOS sensors.} over the years.  
A pixel of CMOS sensor consists of several transistors (typically 4T) to realize charge transfer, reset, row selection, and source follower readout. 
Only one row is switched on when the readout works. An amplifier and an Analog to Digital Converter (ADC) are placed in each column to realise the readout of the whole row signal simultaneously. 
This readout structure enables fast readout speeds of the CMOS sensors, down to several mini-seconds.

Compared to CCD, the CMOS sensor has several other advantages.
The fast readout-speeds equivalently alleviate the requirement for a very low operational temperature---thus the demanding requirement for the cooling---that is needed in order to reduce the influence of the dark current. 
The CMOS is less sensitive to charge transfer degradation caused by charged particles, which are pervading in space. 
The pile-up count rate for single-photon detections is also lifted by 2--3 orders of magnitude. 
The complexity of the readout electronics is reduced, since most of the digital electronics are integrated inside the CMOS sensor. 
Moreover, the cost of the CMOS sensor is usually lower than that of CCDs. 
The CMOS has obvious disadvantages, however.
It has larger dark current non-uniformity than CCD, which is a severe drawback for long-exposure applications. 
There is also non-uniformity in the photon response, arising from the differences of  the amplifiers among pixels.

\begin{figure}[htbp]
\sidecaption[]
\includegraphics[width=0.48\textwidth]{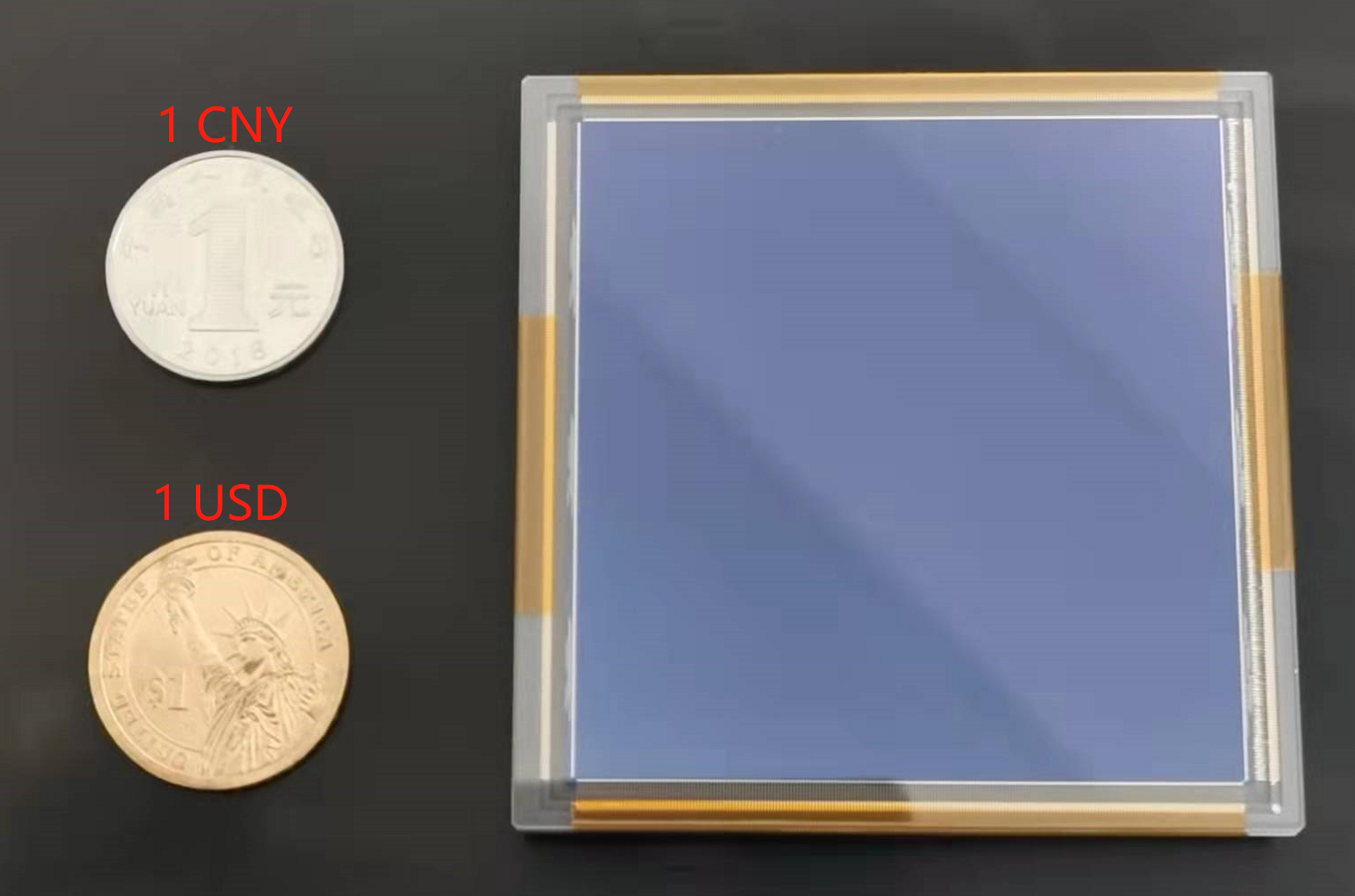}
\caption{Back-illuminated CMOS sensor used for the EP wide-field X-ray telescope, with a size of 6\,cm$\times$6\,cm and a format of 4\,k$\times$4\,k pixels.}
\label{cmos_photo}
\end{figure}

\begin{figure}[htbp]
\sidecaption[b]
\includegraphics[width=0.62\textwidth]{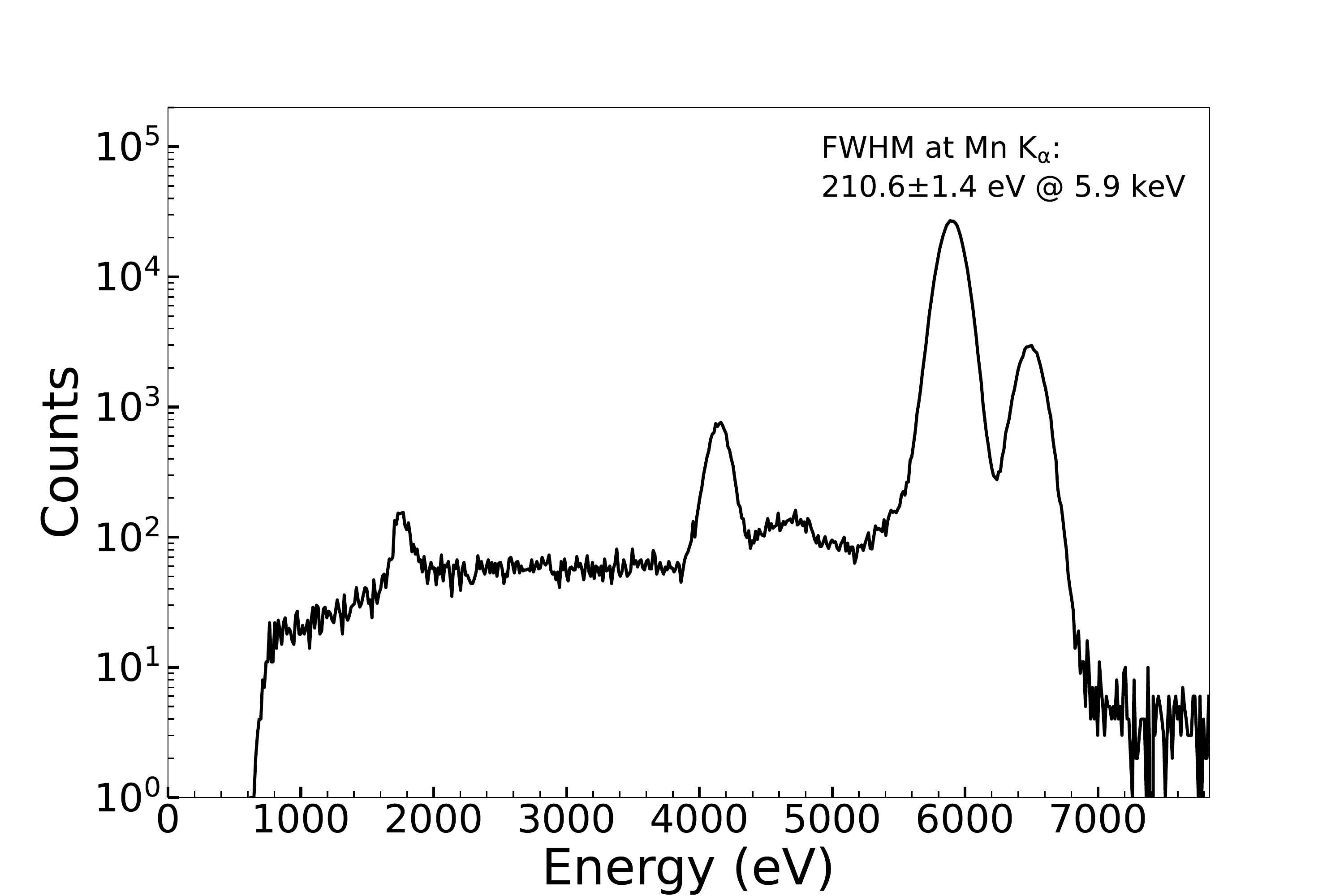}
\caption{X-ray spectrum of an Fe\,55 source obtained at NAOC with the CMOS sensor used for EP.}
\label{fespec}
\end{figure}

A back-illuminated CMOS sensor is adopted and customised for the EP application, considering that soft X-ray photons have short penetration lengths in the detector. 
An example of such CMOS sensors is shown in Fig.~\ref{cmos_photo}, and its specifications are summarised in Table~\ref{tab_cmos}. 
The readout noise is lower than 5~elections at a frame rate of 20~Hz. 
Fig.~\ref{fespec} shows an X-ray spectrum of an Fe\,55 source obtained with the sensor, giving an energy resolution of 210~eV for the K$\alpha$ line at 5.9~keV at a room temperature. 

\begin{table}[]
\caption{Specifications of the EP CMOS sensor}
\label{tab_cmos}     
\begin{tabular}{p{3.5cm}p{4cm}}
\hline\noalign{\smallskip}
Specifications &  value \\
\noalign{\smallskip}\svhline\noalign{\smallskip}
Imaging area & $6\ \rm{cm}\times 6\ \rm{cm}$ \\ 
Pixel size & $15\ \rm{\mu m}\times 15\ \rm{\mu m}$ \\
Epitaxial thickness & $10\ \rm{\mu m}$ \\ 
Number of pixels & $4096\times 4096$ \\ 
Frame rate & 20~fps (max. 100~fps)\\ 
Readout noise & $\rm{< 5.0~e^-}$ at high gain\\ 
Dark current & $\rm{< 0.02~e^-\,pixel^{-1}\,s^{-1}\  @\, -30^{\circ}C}$ \\
& $\rm{< 10~e^-\,pixel^{-1}\,s^{-1}\  @\  20^{\circ}\!C}$ \\ 
Energy resolution & 210~eV@5.9~keV\\ 
                                	&150~eV@1.5~keV\\
Full well capacity & 120,000~e$^-$ \\ 
Power & 1.6~W \\ 
Al coating layer thickness & 200~nm \\ 
\noalign{\smallskip}\hline\noalign{\smallskip} 
\end{tabular}\\
\end{table}

\section{Scientific Instruments}
\label{sec:inst}

The Einstein Probe carries two scientific instruments: a Wide-field X-ray Telescope (WXT) as the monitoring instrument 
and a Follow-up X-ray Telescope (FXT). 
Both telescopes make use of X-ray focusing optics. 

\subsection{Wide-field X-ray Telescope} 
\label{sec:wxt}

\subsubsection{Design of WXT}

\begin{figure}[]
\sidecaption[b]
\includegraphics[width=0.45\textwidth]{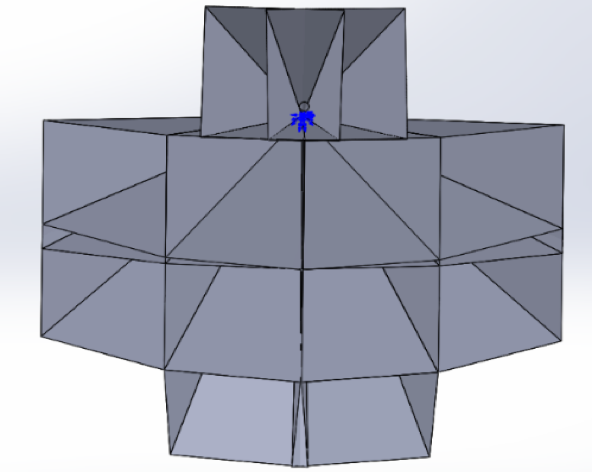}
\caption{Configuration of the WXT FoV, mosaicked by twelve identical modules. The two upper modules have partially overlapping FoVs, which cover the FoV of the FXT (credit: IAMC, CAS).}
\label{fig:wxt-fov}
\end{figure}

To achieve both a wide FoV and X-ray focusing, WXT employs the lobster-eye micro-pore optics.
There are twelve almost identical WXT modules, each of $\sim$300~sq.\,deg. FoV. 
The shape of the overall FoV is shown in Fig.~\ref{fig:wxt-fov},
which makes $\sim 3,600$~sq.\,deg. ($\sim$1.1~steradians) in total and is mostly vignetting-free.
WXT will be the first wide-field X-ray focusing telescope utilising the lobster-eye MPO optics on a large scale. 
The WXT instrument is developed by the CAS institutes\footnote{Jointly developed by the CAS's Shanghai Institute of Technology Physics (SITP) and National Astronomical Observatories (NAOC), with the mirror assemblies developed at NAOC and the optics produced by the NNVT company. Part of the performance tests on the MPO, CMOS and the mirror assembly modules was also provided by ESA as collaboration and carried out at Leicester University and MPE.}.

\begin{figure}[]
\includegraphics[width=0.5\textwidth]{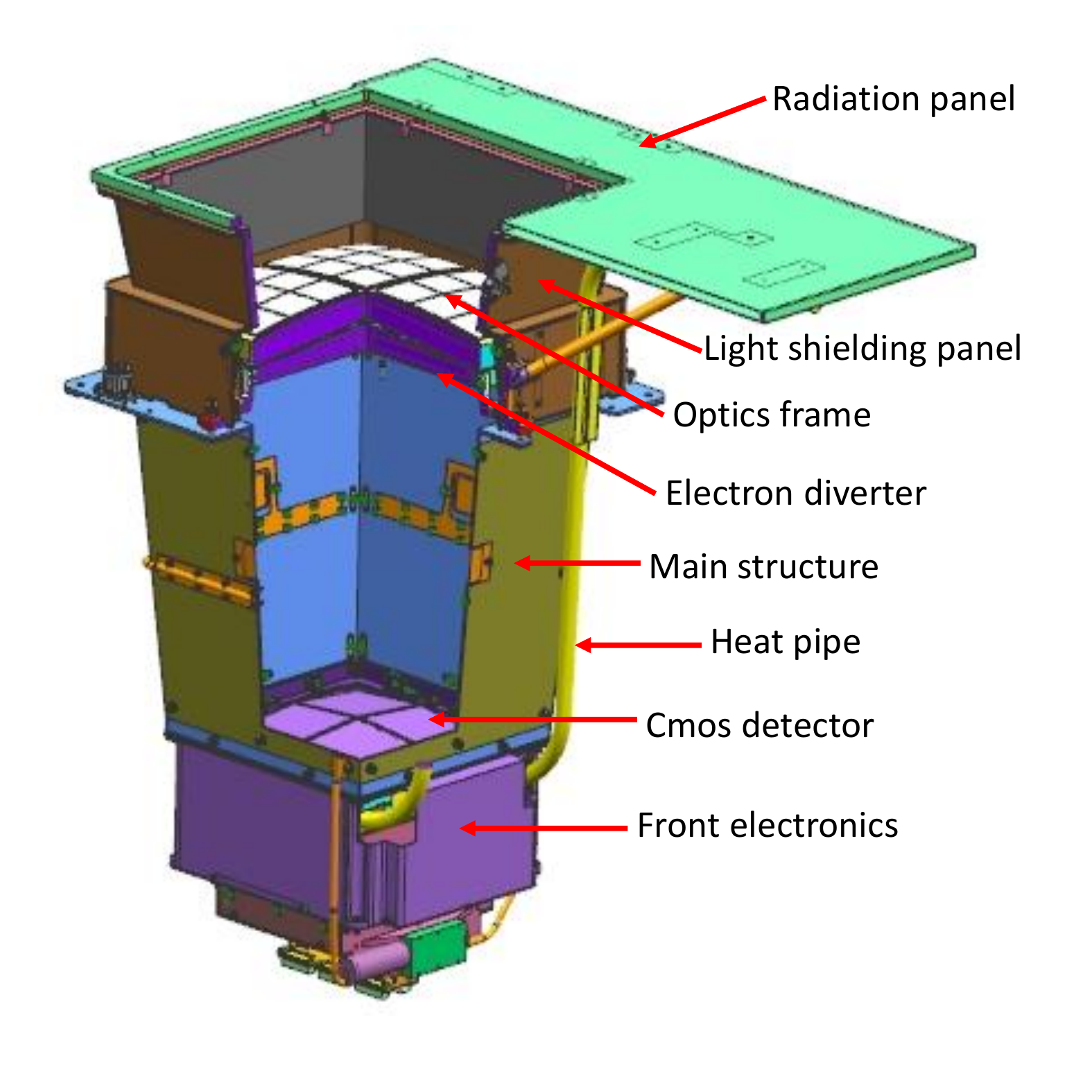}
\includegraphics[width=0.5\textwidth]{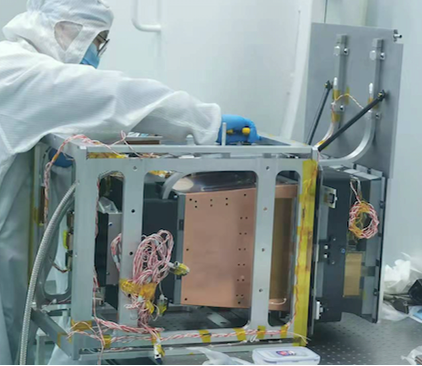}
\caption{Left: design layout of the WXT module (credit: SITP, CAS). Right: one of the qualification models of the WXT module under testing.}
\label{fig:wxt-module}
\end{figure}

Fig.~\ref{fig:wxt-module} (left panel) shows the design layout of the WXT module, whose main components include a MPO mirror assembly, a focal plane detector array and an electronics unit, thermal control and mechanical structure. 
The mirror assembly is the X-ray focusing unit.
An optical baffle (light shielding panel) is attached to the front end of the mirror assembly to shield stray light from the Sun, Moon, and Earth.
A radiation panel is attached to the optical baffle at the top.  
The focal plane detector unit and its front-end electronics are located at the bottom of the WXT module. 
The mechanical support structure is designed to connect the mirror assembly with the detector unit; it is used to shield effectively some of charged particles in orbit .
The main specifications of WXT are listed in Table~\ref{tab:wxt}.

\begin{table}[]
\caption{Specifications of WXT}
\label{tab:wxt}     
\begin{tabular}{p{4cm}p{3cm}}
\hline\noalign{\smallskip}
Specifications &  value \\
\noalign{\smallskip}\svhline\noalign{\smallskip}
Number of modules & 12 \\
Total FoV & $\sim$3,600~sq.deg.  \\   
Focal length & 375~mm\\
Spatial resolution & $\sim$ 5~arcmin (FWHM) \\
Nominal bandpass & 0.5--4.0~keV \\
Energy resolution  & 170~eV  @1\,keV  \\
Effective area$\rm{^a}$  &  2--3~cm$^2$ @1\,keV \\ 
Mass per module & 17~kg \\
\noalign{\smallskip}\hline\noalign{\smallskip} 
\end{tabular}\\
$\rm{^a}$ For the central spot of the PSF only. Note that the effective area varies across the FoV of the WXT module, caused by effects such as the shadowing of the supporting frame and the FoV edges.
\end{table}

The mirror assembly of each module comprises 36 MPO optics mounted on a supporting alloy frame.
A optic has a size of $42.5\times 42.5$~mm$^2$ and a 2.5~mm thickness, with a 375\,mm focal length.
The pores have a size of $40~\mu$m $\times 40~\mu$m and are coated with Iridium to increase reflectivity. 
There are 432 optics in total for the twelve modules.  
An electron diverter is mounted below the optic frame of each module, which will be used to deflect and prevent low-energy electrons in orbit from reaching the detectors.

WXT has a large-format focal plane (sphere) of approximately $420$~mm $\times$ 420~mm in total. 
The focal detector array is composed of four back-illuminated CMOS sensors, each of 6\,cm $\times$ 6\,cm in size.  
The main specifications of the CMOS detector are listed in Table \ref{tab_cmos}. 
A layer of 200~nm aluminum is coated on the surface to block optical light. 
The frame rate used is 20~Hz and the corresponding dead-time is less than 0.05$\%$. 
The readout noise is lower than 5 electrons at 20~Hz. 
The front electronics will read the full-frame image and extract X-ray events as signals that are above a threshold.
All the recorded events are sent to the instrument electronics box for package and then to the satellite and the onboard trigger module for further processing. 
The CMOS detectors will be operated at a temperature of $-30 \pm 2^{\circ}$C, which will be achieved by a thermoelectric cooler. 
A loop heat pipe is used to transfer the heat from the electronics to the main structure of the WXT module.

The electronics control box is used to provide power supply to WXT, packaging the X-ray events data, thermal control, and onboard data processing and triggering (Section~\ref{sec:ot}). 
The front electronics software is designed to be refactorable in orbit.

\begin{figure}[]
\includegraphics[width=0.45\textwidth]{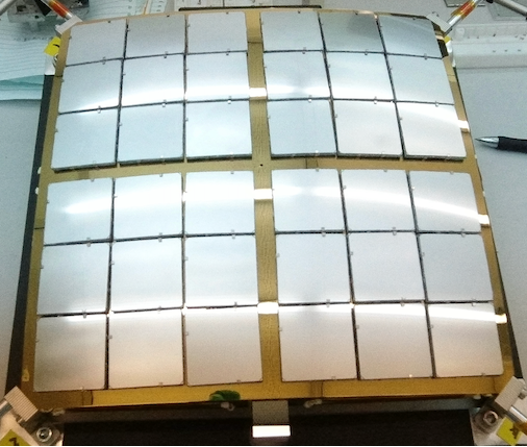}
\includegraphics[width=0.55\textwidth]{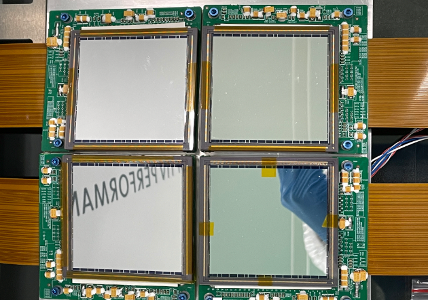}
\caption{Left:  a MPO mirror assembly qualification model for the WXT module developed at NAOC, CAS. Right: a qualification model of the focal-plane detector array for the WXT module developed at SITP, CAS (credit: SITP).}
\label{fig:wxt-ma}
\end{figure}

During phase C of the project, several qualification models of the WXT module were developed with varying engineering status. 
One of these is shown in Fig.~\ref{fig:wxt-module} (right panel), and its mirror assembly and detector array shown respectively in Fig.~\ref{fig:wxt-ma}.
These qualification model modules have undergone extensive engineering and performance tests and calibration. 
The calibration results satisfy most of the design requirements (Zhang et al. in preparation).

\subsubsection{Performance of WXT}

The performance of WXT has been studied extensively via realistic ray-tracing simulations based on the designed WXT model by taking into account the imperfectness of the MPO optic (Zhao et al. 2014, 2017). 
Some parameters were also measured by ground calibration and compared with the simulation results.
The simulated PSF of a MPO optic for WXT is presented in Fig.~\ref{fig:mpo_psf}.

\begin{figure}[]
\sidecaption[b]
\includegraphics[width=0.6\textwidth]{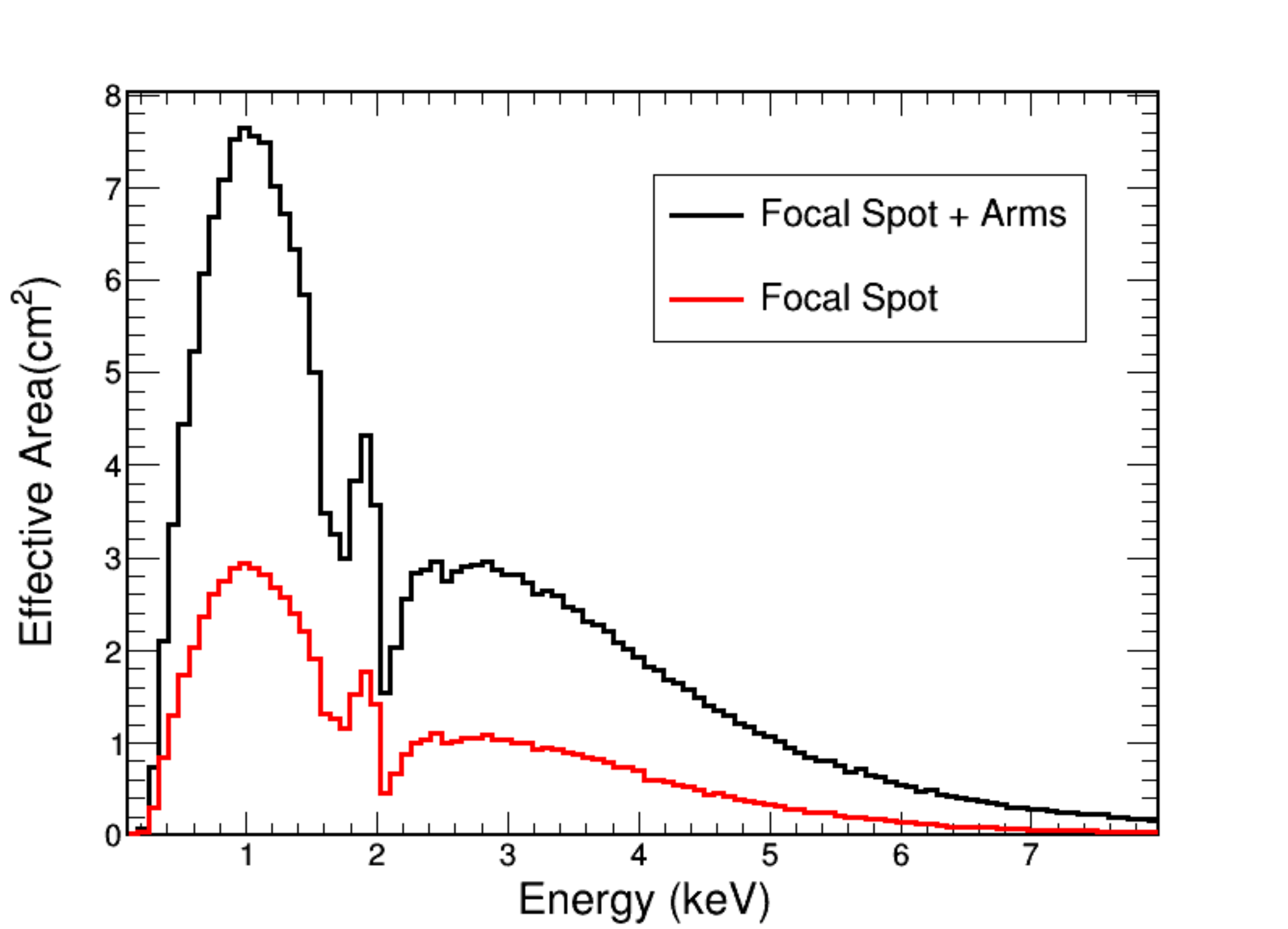}
\caption{Simulated effective area curves of WXT 
for the central focal spot (red) and plus the cruciform arms (black). 
The pores are coated with Iridium, which produces the absorption edge seen at $\sim 2$~keV.
The response of the detectors, coated with a 200\,nm-thick Aluminum layer, is incorporated. 
}
\label{fig:wxt-area}
\end{figure}

The derived energy-dependent effective areas of WXT are shown in Fig.~\ref{fig:wxt-area}. 
The effective area for the central spot peaks around $\sim$3~cm$^2$ at 1~keV.  
When all the photons focused onto both the central spot and the two cross-arms are collected, the effective area increases to $\sim$7.6~cm$^2$ at 1~keV. 
The areas drop towards both the higher and lower energies, due mainly to the decreasing X-ray reflectivity with increasing energy, the stronger absorption at lower energies, and the energy dependence of the detector quantum efficiency. 
The effective area curves define basically the detecting bandpass of WXT.
The nominal band is from 0.5 to 4.0~keV, beyond which the effective area drops rapidly.
In fact, the effective area also varies across the FoV of the WXT module.
This is caused mainly by effects such as the shadowing of the supporting frame and being close to the edges of the FoV.

\begin{figure}[]
\sidecaption[b]
\includegraphics[width=0.58\textwidth]{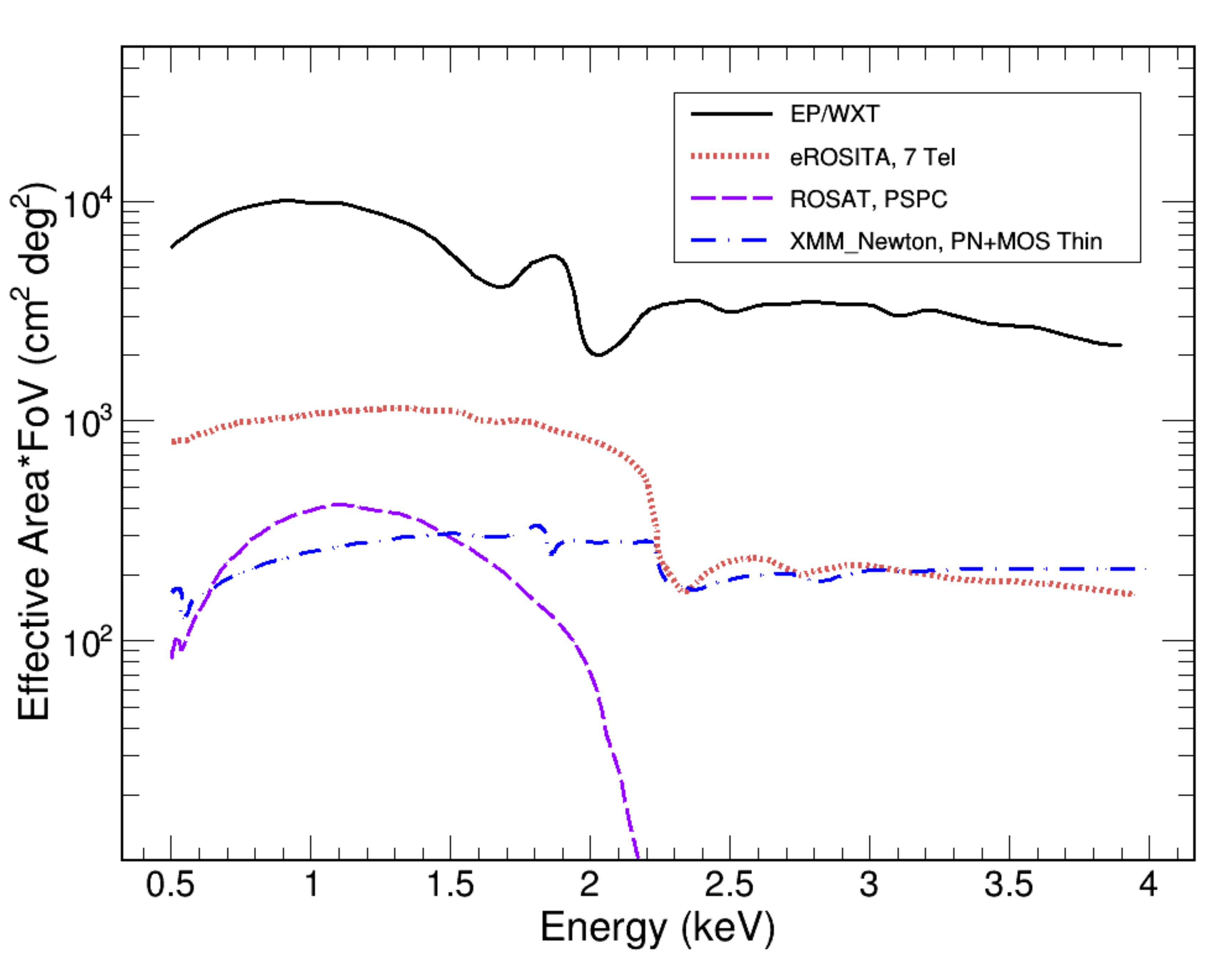}
\caption{Representative grasp (effective area times FoV) of WXT as a function of photon energy (black).  As a comparison, the grasp parameters of several X-ray focusing telescopes are overplotted (update based on the results in Zhao et al. 2014).}
\label{fig:wxt-grasp}
\end{figure}

Although the effective area for most directions is only the order of several cm$^2$, its variations are small across almost the entire FoV.
This leads to an enormous grasp---effective area times FoV, peaking around $10^4$~cm$^2$\,(sq.\,deg.) at 1~keV.
As shown in Fig.~\ref{fig:wxt-grasp}, WXT has the largest grasp in soft X-rays among the X-ray focusing telescopes built so far.

\begin{figure}[]
\sidecaption[b]
\includegraphics[width=0.5\textwidth]{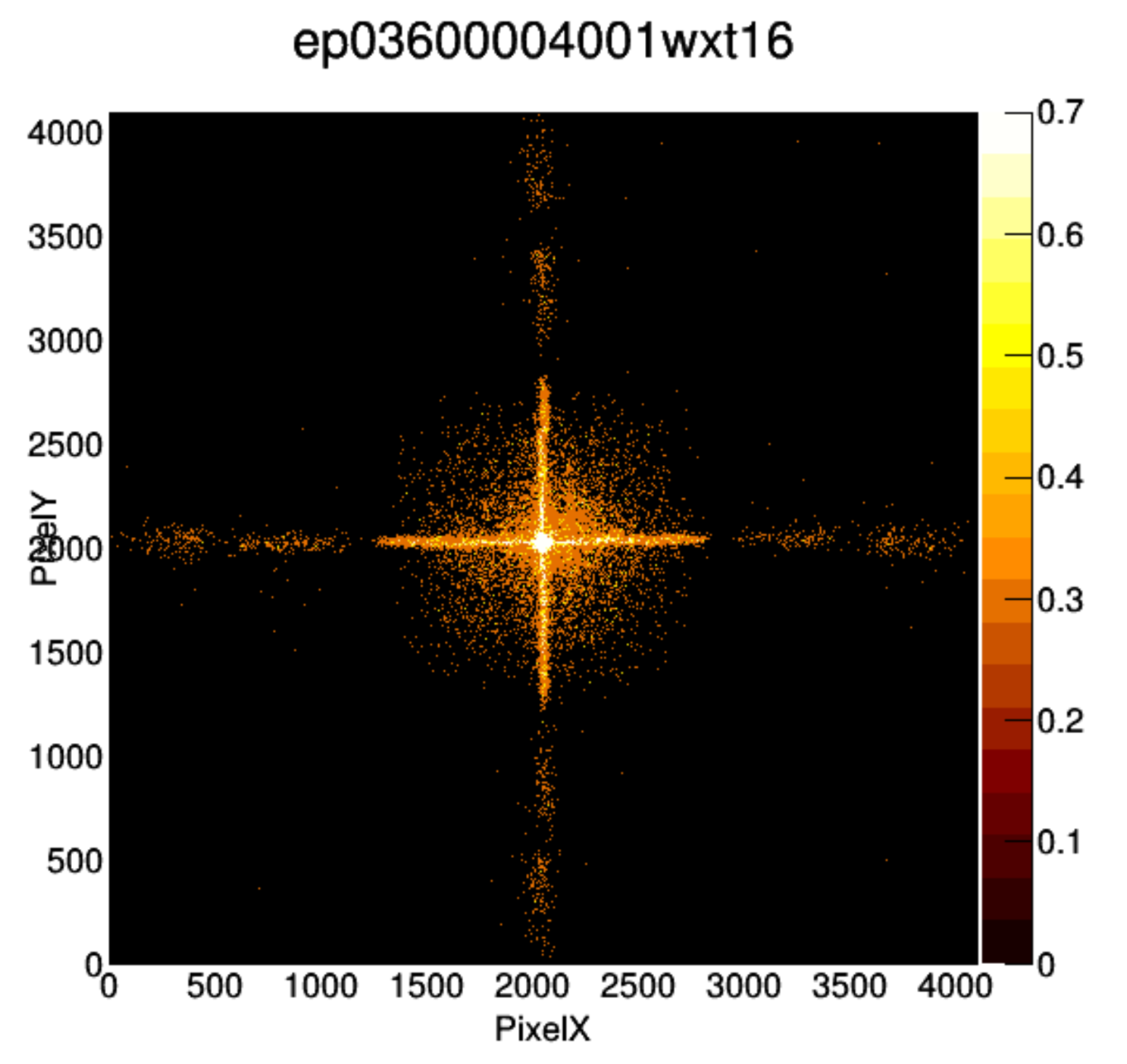}
\caption{True X-ray image produced by one mirror assembly qualification model of the WXT module developed at NAOC (measured at MPE/Panter by using a parallel beam of a Cu-L$\alpha$ line source of 930~eV). 
The PSF is symmetric, with an angular resolution $<$5~arcmin in FWHM for the central spot (Courtesy ESA/MPE).}
\label{fig:wxt-PSF}
\end{figure}

Each of the MPO optics was tested in X-rays, and the majority have an angular resolution of $\sim 5$~arcmin (FWHM) or better.
Two qualification models of the WXT module have been tested and calibrated with X-ray beams. 
Good agreements on both the PSF and effective area were found between the measurements and simulations. 
In most of the directions across the 300~sq.\,deg.\ FoV of the WXT module,  the angular resolution was found to be $\le 5$~arcmin (FWHM, central spot) on average, and the effective area in the range of 2--3~cm$^2$. 
Fig.~\ref{fig:wxt-PSF} displays an X-ray image produced by one of the WXT mirror assemblies, which was obtained at MPE/Panter by using a parallel beam of a Cu-L$\alpha$ line source of 930~eV.

The EP satellite will be subject to a complex space environment full of cosmic rays including photons and various types of energetic charged particles. 
Simulations of the WXT background (Zhao et al. 2018) were carried out by using the toolkit of Geant4 (Agostinelli et al. 2003), as well as
XRTG4 (Buis et al. 2009) to simulate the grazing incidence of X-rays.
The X-ray background is composed of the cosmic X-ray background (CXB) and the diffuse soft X-rays within the Galaxy, dominating below 1~keV.
Different from the X-ray background whose contribution is mainly determined by the optics of WXT, the background produced by energetic charged particles depends strongly on the thickness of the depletion-layer of the detector.
As a result, in the 0.5--4 keV band, a background level of about 0.3~counts~s$^{-1}$\,cm$^{-2}$ is predicted for the WXT detectors. 
Among them, the diffuse X-ray background contributes about 0.2~counts~s$^{-1}$\,cm$^{-2}$ and dominates the soft X-ray band below 2~keV (Zhao et al. 2018).

The detection sensitivity of WXT is determined by a number of factors, mainly the effective area and PSF of the entire imaging system, the background count rate, the energy bandpass, and the source spectral shape. 
By using the values derived from the simulations discussed above as well as the PSF  measured from the ground calibration of the qualification models, the detection sensitivity was evaluated from simulations. 
For a point-like source, the medium values of the limiting flux in the 0.5-4\,keV band are approximately 
$8.9 \times10^{-10}$\,ergs\,s$^{-1}$\,cm$^{-2}$  (27.6\,mCrab) for an exposure of 10~seconds, 
$1.2 \times10^{-10}$\,ergs\,s$^{-1}$\,cm$^{-2}$   (3.9\,mCrab) for 100\,s, and 
$0.26 \times10^{-10}$\,ergs\,s$^{-1}$\,cm$^{-2}$  (0.8\,mCrab) for 1,000\,s
(assuming a power law spectrum with a photon index -2 and a Galactic absorption column  $3\times10^{20}$\,cm$^{-2}$). 
It is clear that EP WXT is much more sensitive compared to the previous and current wide-field X-ray monitors by one order of magnitude or more.
However, these values should be considered as nominal, as the true sensitivity depends on the in-orbit performance of the instrument and strongly on the background level determined by the actual environment of charged particles.

Because of the fast frame rate of the CMOS detectors, the pileup fraction is predicted to be lower than 1$\%$ for a source as bright as 10~Crab.

\subsection{Follow-up X-ray Telescope}

The Follow-up X-ray Telescope (Chen et al. 2020) is designed to perform prompt X-ray follow-up observations of faint transients and variable sources detected with WXT.
The main requirements are to locate sources down to an accuracy of several arcseconds  to improve upon the WXT source positions ($\sim$1~arcmin), and to measure the source X-ray spectra, light curves, and images.
It may also be used for observations of targets of opportunity (ToO) of interest. 
The development of FXT is an international collaborative effort\footnote{The development of FXT is  led by the Institute of High-Energy Physics (IHEP), CAS, with significant contributions from ESA and MPE.}.
 
\subsubsection{Design of FXT}

\begin{figure}[b]
\centering
\includegraphics[width=9cm, angle=0]{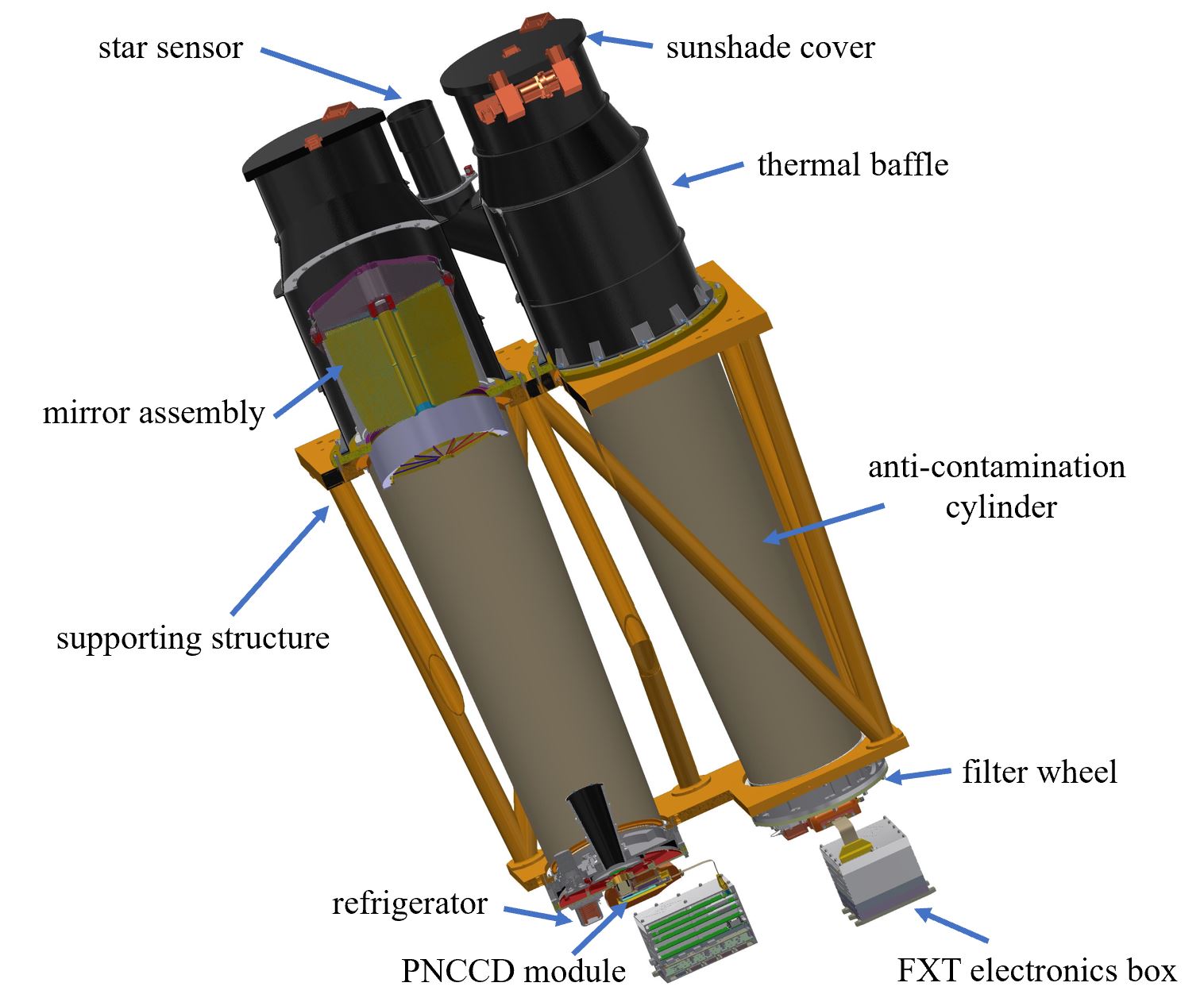}
\caption{Configuration and components of the two-telescope units of FXT.}
\label{fig:FXT-config}
\end{figure}

In order to achieve the science objectives and requirements of the mission, it is required for FXT to have a large effective area, good angular resolution, and capabilities of timing, imaging and spectroscopic observations.
A  Wolter-I nested telescope is adopted.
The main specifications and designed performance are listed in Table~\ref{tab:FXT}.
FXT is composed of two telescope units, FXT-A and FXT-B, which have almost the same  structural settings and are co-aligned (Fig.\,\ref{fig:FXT-config}). 
Compared to a single-telescope design, the two-telescope units not only  improve greatly the effective area, but also enhance the redundancy and reliability.  
The observation modes of the two units are independent and can be set flexibly, which include the full-frame, partial-window, and timing modes (Table~\ref{tab:ccdmode}). 
By choosing the different modes, FXT can observe sources with a large dynamic range of fluxes.

\begin{figure}
\centering
\includegraphics[width=0.45\textwidth]{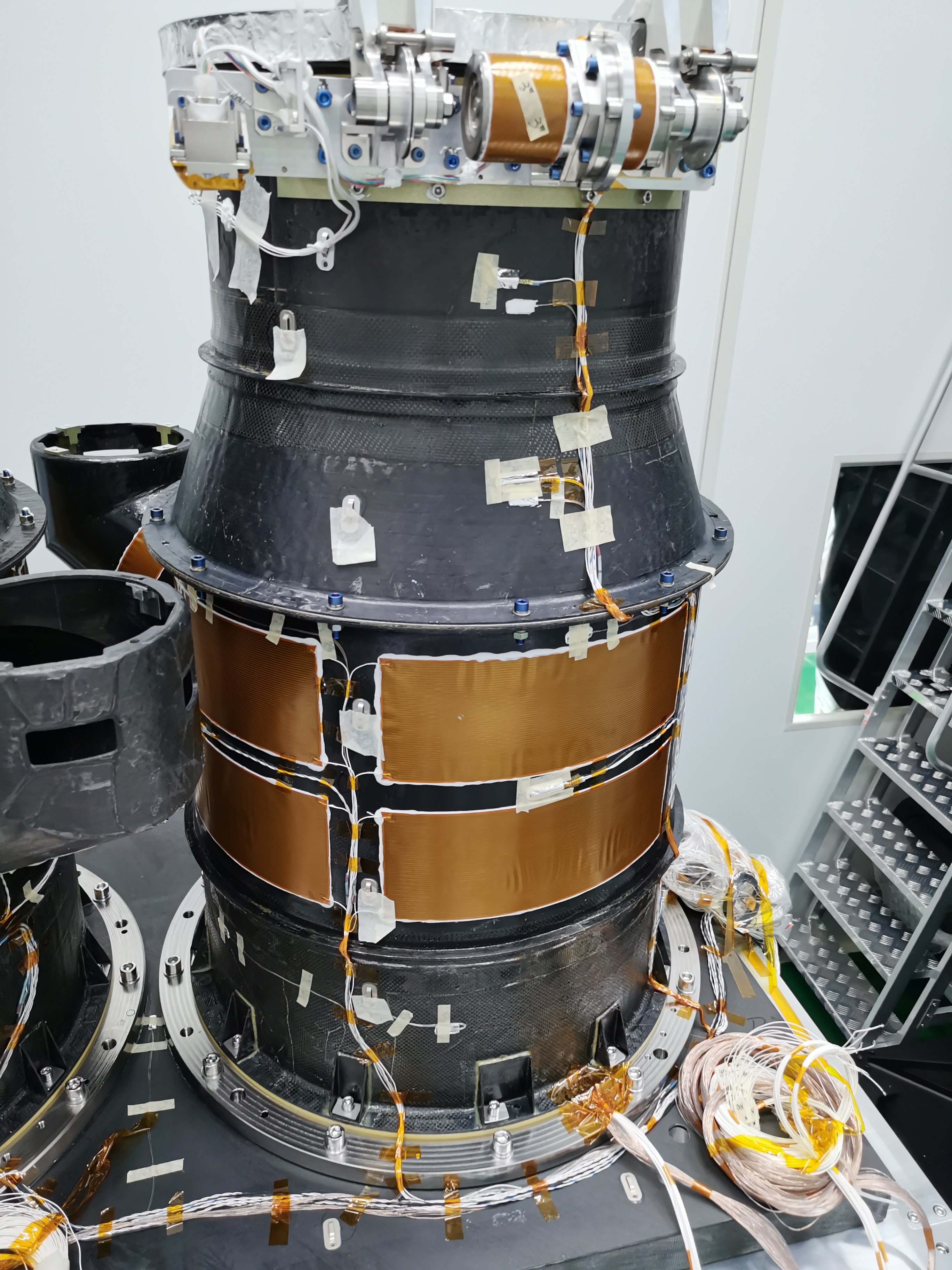}
\includegraphics[width=0.45\textwidth]{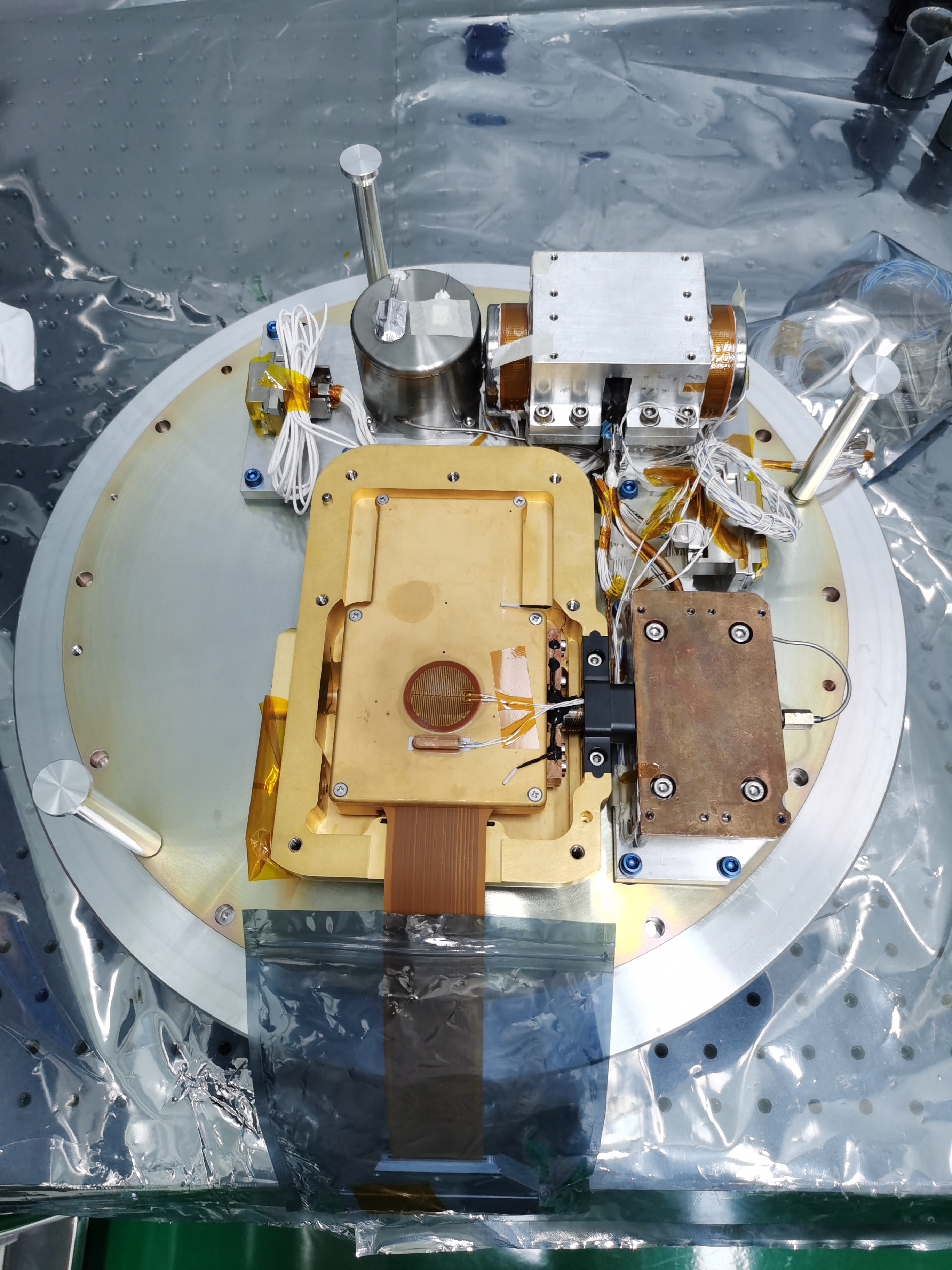}
\caption{Qualification model of the upper composite (left) and the lower composite (right) of FXT developed at IHEP, CAS.}
\label{fig:upper}
\end{figure}

Each FXT unit is composed of three parts, an upper composite (Fig.~\ref{fig:upper}, left panel), a lower composite (Fig.~\ref{fig:upper}, right panel), and supporting structure. 
The upper composite is composed mainly of a mirror assembly, a thermal baffle, a sunshade cover, and an optical reference flange. 
A star sensor is also set on the shading cylinder. The lower composite includes a filter wheel, a detector unit and a refrigerator.

\begin{table}
\caption{\label{tab:FXT} Specifications and goal performance of FXT}
\hspace{0.2cm}
\smallskip
\begin{tabular}{p{4cm}p{3.5cm}}
\hline
Specifications           &  Value         \\
\noalign{\smallskip}\svhline\noalign{\smallskip}
Number of units &  2 \\
Focal length & 1.6~m \\
No. mirror shells (per unit) & 54 \\
Field of View      & $1^\circ \times 1 ^\circ$ \\
Effective area (2 units) & 600~cm$^2$ @ 1.25 keV  \\
Angular resolution & 30~arcsec (HPD)                  \\
Energy resolution  &  120~eV @ 1.25~keV (FWHM)     \\
Energy range & 0.3 -- 10~keV  \\
Source locating uncertainty & 4~arcsec (1 $\sigma$)\\
\hline
\end{tabular}
\end{table}

As a follow-up telescope, FXT should have an appropriate dimension and weight to facilitate the configuration of the satellite platform. 
A Wolter-I telescope with a short focal length has the advantages of a small volume and light weight. 
A design of the mirror assembly similar to that of the eROSITA telescopes (Predehl et al. 2021) was adopted, which has a focal length of 1.6~m.
The two mirror assemblies of FXT-A and FXT-B are provided by ESA and MPE, respectively, as contributions to the mission. 
For the focal plane detectors, the PNCCD detector module (Meidinger et al. 2006) that is basically the same as that used for eROSITA was adopted, which was developed and provided by MPE.

The FXT mirror assembly,  manufactured at Media Lario in Italy, consists of 54 nested gold-coated nickel shells.
An X-ray baffle (Friedrich et al. 2014) is installed above the mirror assembly, which is designed not only to limit stray light, but also to reduce the temperature fluctuations of the mirror caused by variations of the thermal environment. 
An electron diverter is installed under the mirror assembly to deflect low-energy electrons in orbit to prevent them from reaching the detector, so as to help reduce the detector background.

\begin{figure}
\sidecaption[b]
\includegraphics[width=0.55\textwidth]{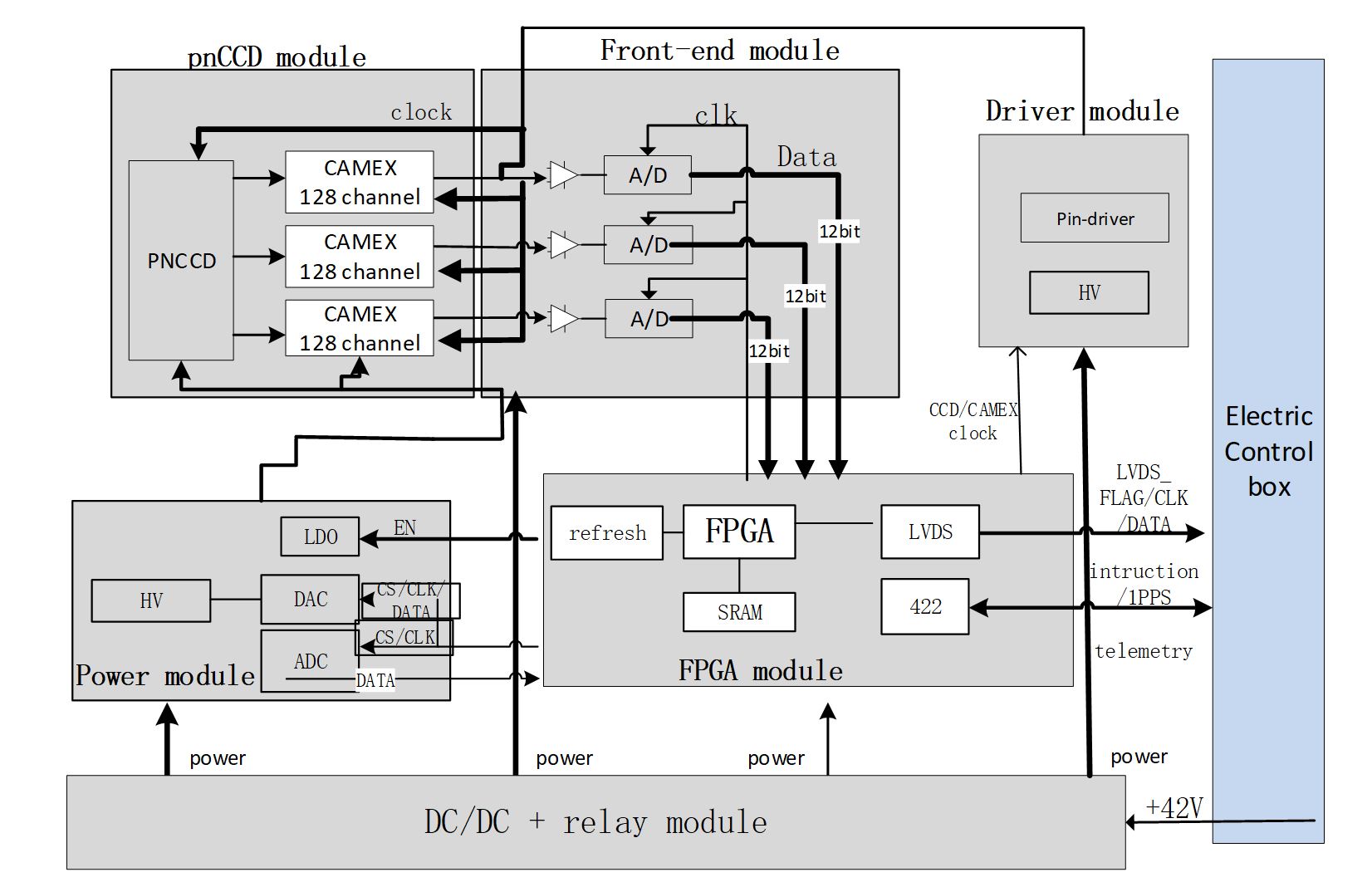}
\caption{Block diagram of the FXT electronics.}
\label{fig:electronics}
\end{figure}

The PNCCD is an advanced frame-transfer, back-illuminated and fully depleted CCD sensor.
There are 384 $\times$ 384 pixels, each of 75 $\times$ 75 microns. 
The PNCCD has a good quantum efficiency in the energy range of 0.3--10~keV. 
The detector module is composed of PNCCD chips, CAMEX, and related electronics circuits, which are packaged on a ceramic plate. 
It is connected to an electronic box (developed by IHEP) through flexible circuit boards. The electronics box is mainly responsible for driving and data acquisition of the PNCCD. Its circuit block diagram is shown in Fig.~\ref{fig:electronics}.
The PNCCD can operate in three different readout modes, 
the full-frame, partial-window and timing modes, by changing the readout size or dimension of the imaging area (Table~\ref{tab:ccdmode}). 

\begin{table}
\centering
\caption{\label{tab:ccdmode} Operation modes of PNCCD}
\hspace{0.2cm}
\smallskip
\begin{tabular}{p{2cm}p{2cm}p{2cm}p{3cm}}
\hline
Mode     &  Pixel region                & Readout time & Bright source limit    \\
         & (row $\times$ column)  &       & (pile-up fraction $<10\%)$\\
\noalign{\smallskip}\svhline\noalign{\smallskip}
Full frame      &  384 $\times$ 384  & 50 ms    & 10 mCrab    \\
Partial-window  &  61 $\times$ 128  &  2 ms    & 200 mCrab    \\
Timing          &  384 $\times$ 128  &  23.6 ${\rm {\mu}}$s/row    &  5 Crab    \\
\hline
\end{tabular}
\end{table}

The optimal working temperature of the FXT CCD is $-90^\circ$C. 
A helium pulse tube refrigerator was adopted for each FXT unit. 
The refrigerator is developed at the Technical Institute of Physics and Chemistry, CAS. 
Testing experiments using a qualification model of the refrigerator demonstrated that the detector temperature can be stabilised at $-90^\circ$C $\pm$ 0.5$^\circ$C.

In order to adapt to various observing conditions, the filter wheel is equipped with six operating positions: thin filter, medium filter, open, closed, hole, and calibration position. The specific settings are shown in Table~\ref{tab:wheel}. 

\begin{table}
\caption{\label{tab:wheel} Configuration of the FXT filter wheel}
\hspace{0.2cm}
\smallskip
\begin{tabular}{p{3cm}p{4cm}p{2cm}}
\hline
Position      & Material (thickness) & Diameter\\
& & (mm) \\
\noalign{\smallskip}\svhline\noalign{\smallskip}
Thin filter   & PI (200 nm) + Al (80 nm) & 50 \\
Medium filter & PI (400 nm) + Al (200nm) & 50 \\
Open position &  -                       & 50 \\
Closed position &  Al(2 mm)    & 50 \\
Hole position & PI (200 nm) + Al (80 nm) & 22 \\
Calibration position & $^{55}$Fe radioactive source  & - \\ 
\hline
\end{tabular}
\end{table}

\subsubsection{Performance of FXT}

\begin{figure}
\sidecaption[b]
\includegraphics[width=0.62\textwidth]{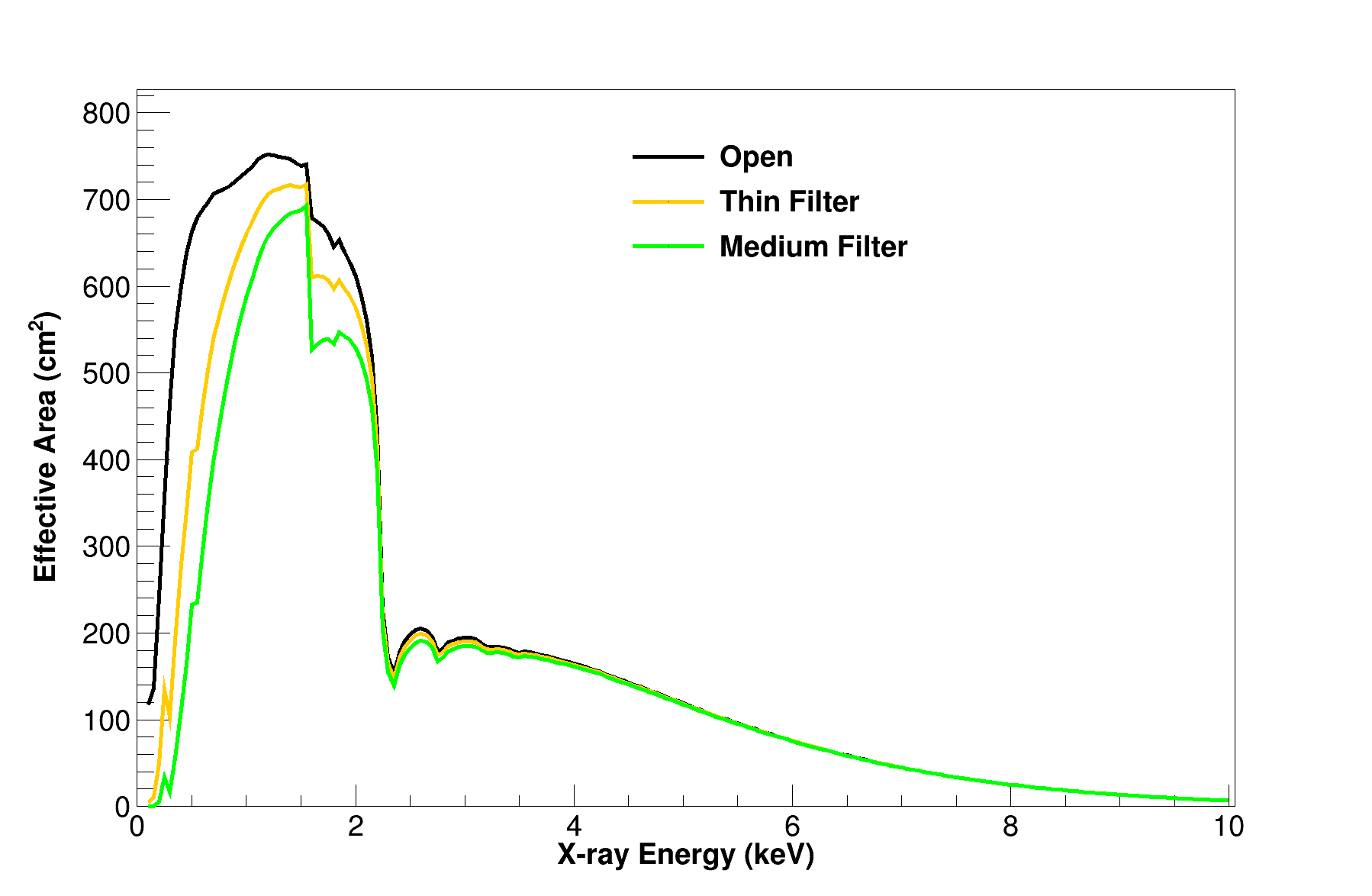}
\caption{Simulated effective area of the combined two FXT units. Effects of the detector quantum efficiency and the filters are taken into account.}
\label{fig:eff}
\end{figure}

The effective area of the mirror assembly was estimated by performing simulations, which is shown in Fig.~\ref{fig:eff} for the combined two FXT units.
The combined effective area reaches $\sim 700$~cm$^2$ at around 1~keV, which is excellent for a dedicated follow-up telescope. 
The combined effective area  is $94$~cm$^2$ at the Fe K line energies around 6~keV, comparable to that of the Swift/XRT.

The background of FXT was simulated by using Geant-4. The effects of various background components in space were considered, including the cosmic X-ray background,  primary and secondary cosmic-ray particles, albedo gamma-rays, and low-energy protons near the geomagnetic equator, etc. (Zhang et al. 2022). 
Among them, the CXB within the FoV dominates the background below 2~keV. 
For particle-induced background, equatorial low-energy protons account for the largest proportion. 
The particle background on the detector is about 0.03~counts\,s$^{-1}$\,keV$^{-1}$ in total. 

\begin{figure}
\sidecaption[b]
\includegraphics[width=0.55\textwidth]{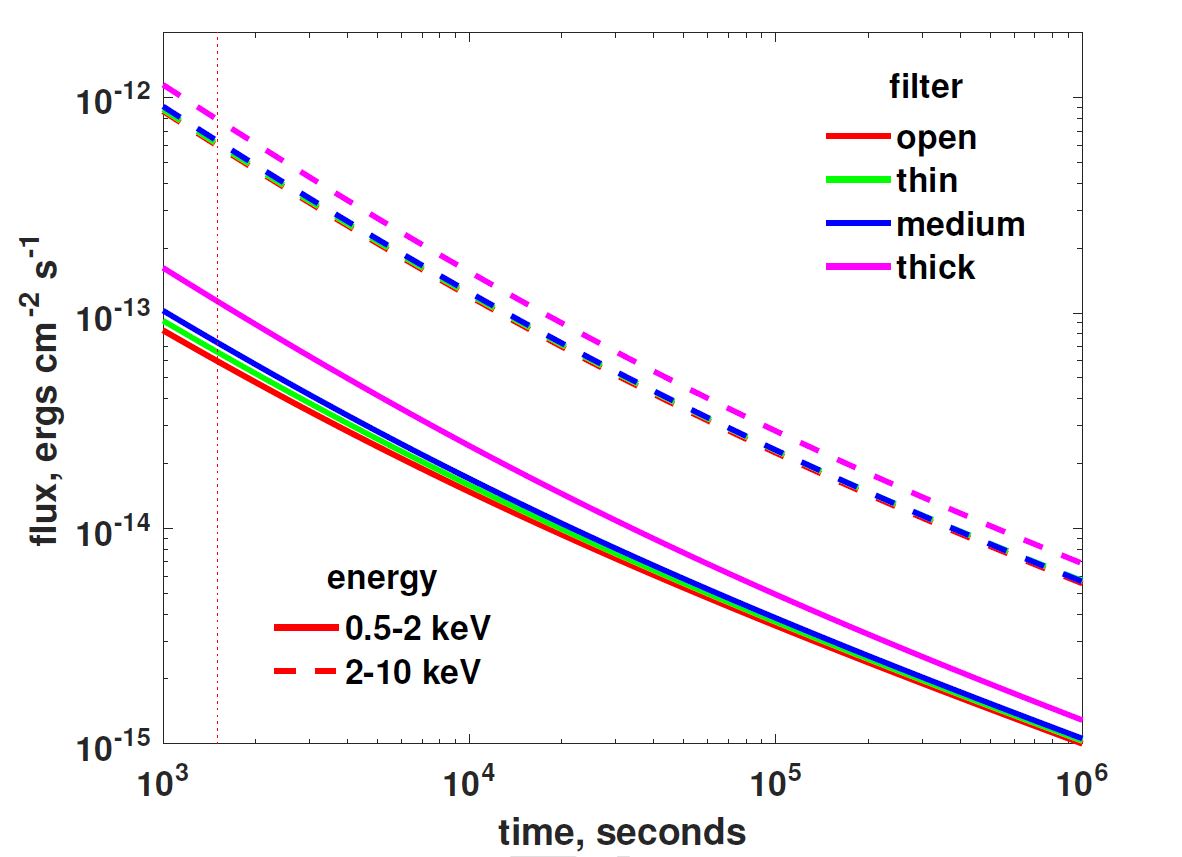}
\caption{Simulated sensitivity of FXT for one telescope unit in the 0.5--2~keV and 2--10~keV bands for different filter settings (figure from Zhang et al. 2022).}
\label{fig:sens}
\end{figure}

Based on the simulated background level, effective area, and designed PSF, the detection sensitivity for a point-like source was estimated by simulations. 
The results show that the sensitivity for one telescope unit can achieve the order of 10$^{-14}$~ergs\,s$^{-1}$\,cm$^{-2}$ in 0.5--2 keV with a 25-minute exposure. 
The sensitivity for one telescope unit is shown in Fig.~\ref{fig:sens} as a function of exposure (Zhang et al. 2022).

\begin{figure}
\includegraphics[width=0.60\textwidth]{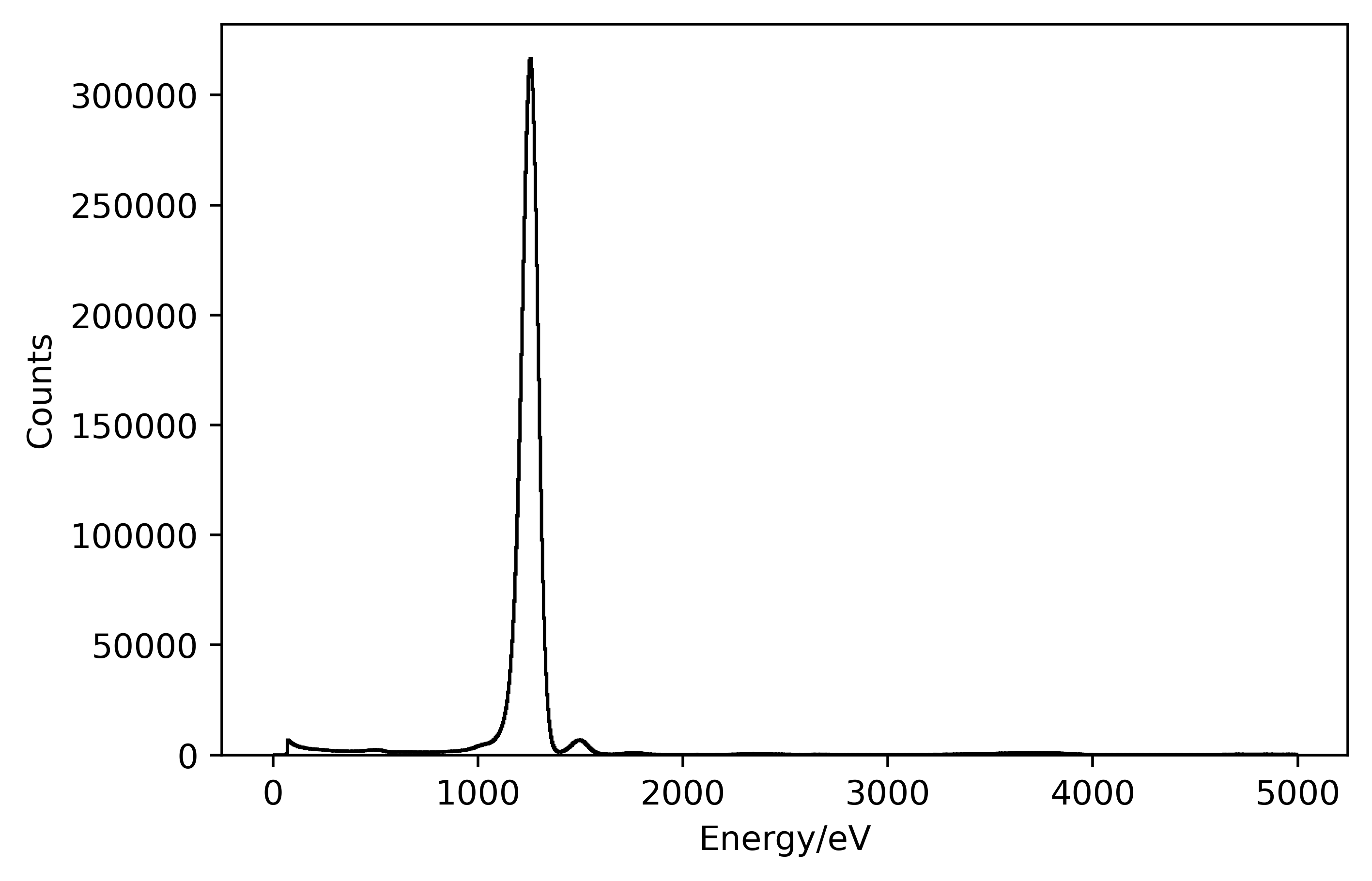}
\includegraphics[width=0.39\textwidth]{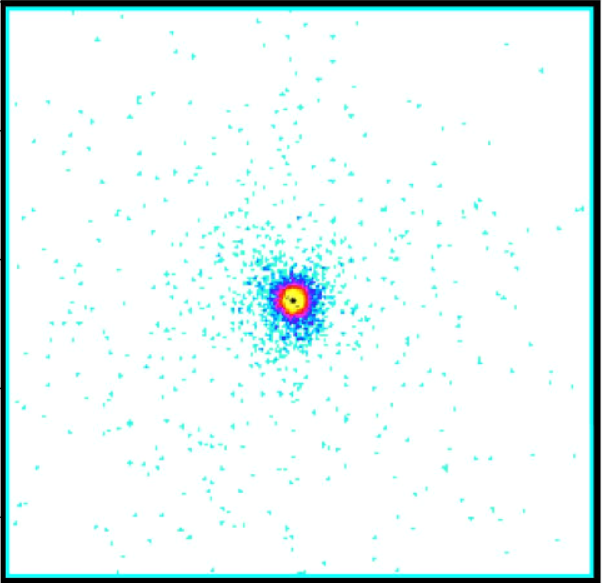}
\caption{Left: a measured PNCCD spectrum of the Mg X-ray fluorescence line obtained at IHEP. Right: an X-ray image of the PSF of the FXT mirror assembly qualification model, showing an angular resolution of 21" in HPD measured at 1.49~keV (courtesy ESA/MPE).}
\label{fig:Mg}
\end{figure}

The PNCCD modules have been extensively tested at IHEP. 
The energy resolution can reach 97~eV at 1.25~keV (Fig.\,\ref{fig:Mg}, left panel), and the overall equivalent noise of the system is about 3.5~e$^{-}$ on average. 

A qualification model of the mirror assembly has been tested and calibrated at MPE/Panter, and will be further tested at the 100m calibration facility at IHEP. 
The measured spatial resolution, derived from the PSF image (Fig.~\ref{fig:Mg}, right panel), is 21".0$\pm 0.3$  in HPD (half power diameter) and the effective area is $367\pm$ 6 cm$^2$, both measured at 1.49~keV using the Al K line. 

\section{Satellite and mission profile}
\label{sec:miss}

The EP project comprises the following major systems: the spacecraft and payload, launcher and launch centre, tele-command control, mission centre, and science centre. 
The project is currently in the flight model development phase. 
The mission lifetime is three years as design and five years as goal.
The satellite is to be launched with a Long March 2-C rocket from the Xichang Satellite Launch Centre.  
The satellite will adopt a low-earth circular orbit with an altitude of about 600~km (a period of $\sim$97~minutes) and an inclination of 29~degrees. 
This section describes the satellite and the overall mission profile, focusing on the components closely related to the science operation and performance.

In the designs of the satellite system and ground segment, the requirements for quick response were considered;
that is, prompt onboard/ground follow-up observations and fast satellite-ground telecommunications in two ways.

\subsection{Satellite system} 

The EP satellite system consists of the scientific payloads WXT and FXT, and a spacecraft that is especially designed to accommodate and serve the payloads and their operations. 
An onboard data processing and triggering system is also designed (Section~\ref{sec:ot}). 


\begin{figure}[t]
\includegraphics[width=1.0\textwidth]{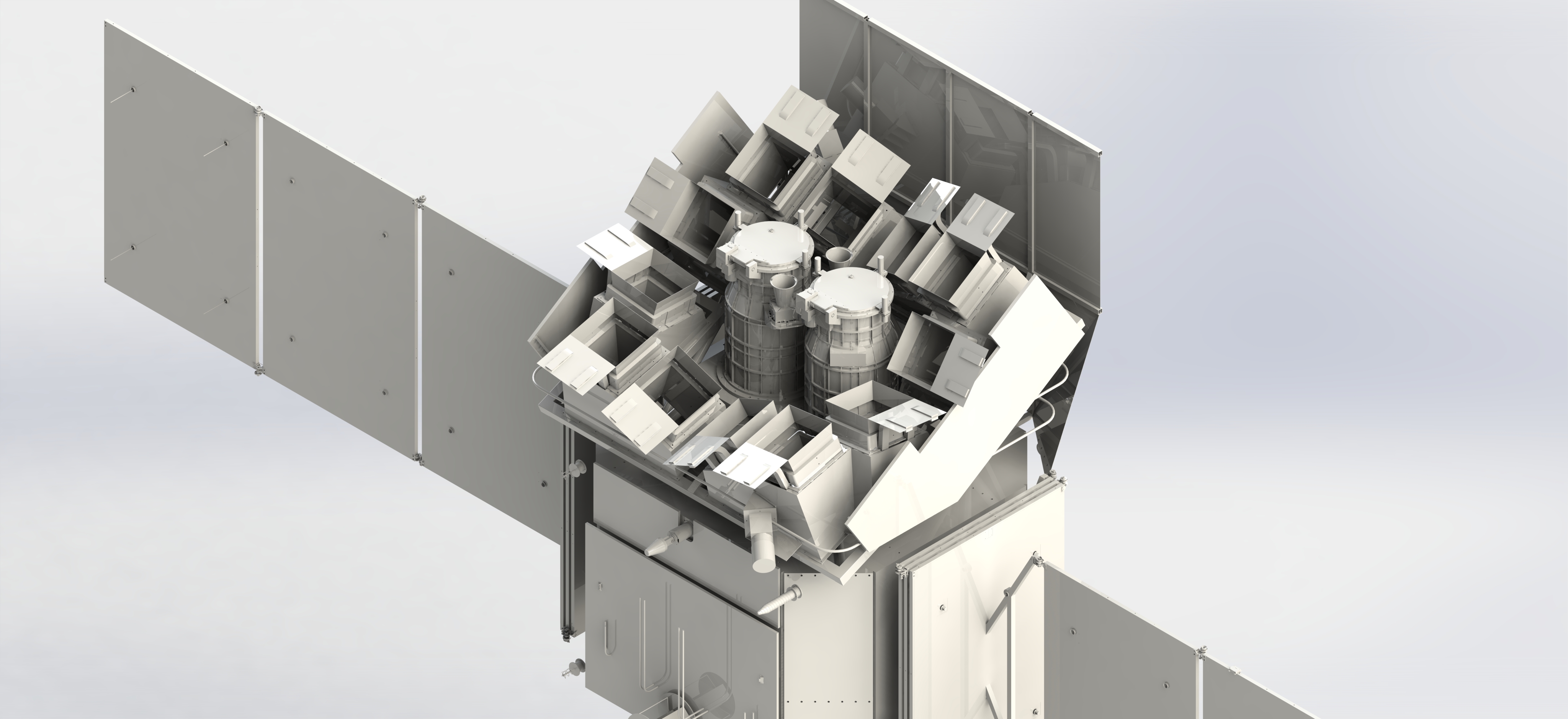}
\caption{Layout of the payloads and spacecraft of the Einstein Probe satellite. The two FXT units are placed at the centre, surrounded by the twelve WXT modules (credit: IAMC, CAS)}
\label{fig:satellite}       
\end{figure}

\begin{figure}[t]
\sidecaption[b]
\includegraphics[width=0.48\textwidth]{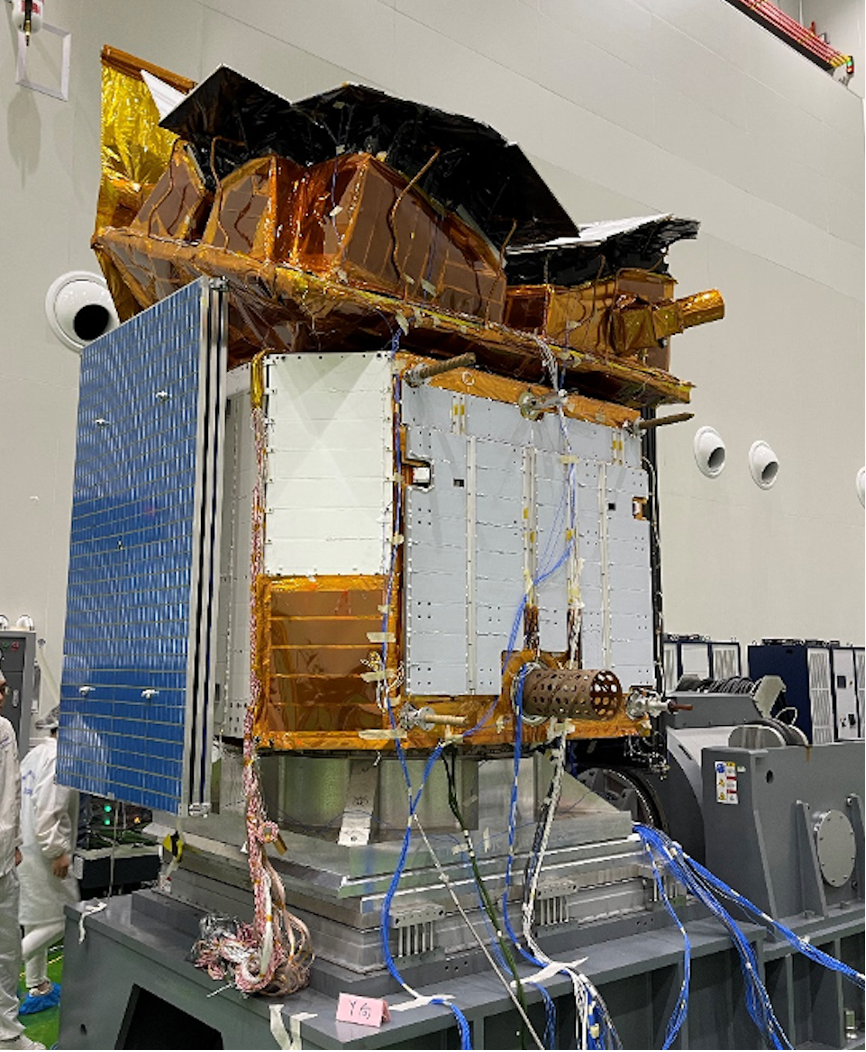}
\caption{EP satellite qualification model in preparation for mechanical tests  (credit: IAMC, CAS).}
\label{fig:sat-qm}       
\end{figure}

The spacecraft\footnote{Developed by the Innovative Academy for Micro-satellites of CAS (IAMC).} is composed of the following subsystems:  structure, thermal control, attitude and orbit control, power supply and overall circuits, on-board computer, telemetry and tele-control, communication for alert and ToO messages. 
The spacecraft is equipped with six momentum wheels to enable quick manoeuvre, at a slew speed of 15~degrees per minute. 
The power system is composed of solar-panels, lithium batteries and power controller. 
All spacecraft commanding and controlling tasks, including attitude control and FDIR (failure detection, isolation and recovery) is performed by the Satellite Management Unit (SMU). 
All the payloads are managed by the payload data processing unit (PDPU). 
Payload science data are routed to the solid-state mass memory. 
%
The spacecraft is designed to achieve the following performance of attitude control: 
\begin{itemize}
\item{Attitude control: three-axis stabilising, inertial orientation during observation}
\item{Manoeuvring speed: better than 60~degrees in 4~minutes}
\item{Pointing accuracy: better than 0.05~degrees}
\item{Stability: better than 0.0005~degrees\,s$^{-1}$}
\end{itemize}
The payload weights 564~kg, and the satellite has a weight $\le$1,450~kg and an average power 1,212~W. 
In phase C, a satellite qualification model (Fig.~\ref{fig:sat-qm}) has been developed at IAMC, which has undergone all kinds of the required tests including vibration and acoustic tests, thermal balancing thermal vacuum tests, EMC, and solar-panel deployment test.

\subsection{Onboard data processing and triggering}
\label{sec:ot}

One system requirement of the mission is real-time detection of transients on the fly and quick response for both onboard and ground-based follow-ups.
To this end, an onboard data processing and triggering system (ODPTS) was designed. 
While data acquisition is ongoing during a WXT observation, the onboard computer will search for transients in real time by processing the acquired WXT data.
The search is performed over each CMOS chip and over a range of timescales. 
First, two 1D positional histograms of the detected photons, projected onto the horizontal and vertical directions respectively, are continuously accumulated. 
Potential source candidates are identified from the two 1D histograms, and their positions are recorded and matched to yield the possible 2D positions. 
Then, photons around these 2D positions are extracted for evaluation and screening, yielding a list of detected sources. 
The sources are further compared against a source catalogue to find any transients.
Whenever a new transient is picked up, the ODPTS will trigger a FXT observation. 
Meanwhile, an alert message will be downlinked to ground in time (Section~\ref{sect:com}).
Transient alerts will be made public upon receipt and manual screening in real time, to trigger follow-up observations from the worldwide community.

It has been demonstrated that the data processing for all the 48 CMOS detectors can be achieved within 10~seconds using the real hardware. 
A source detection sensitivity of $\sim$1~mCrab for a typical 1,000~s exposure of one survey pointing can be achieved.
The source location accuracy is about 2~arcmins, as demonstrated by both simulations and ground experimental tests.


\subsection{Science operation}
\label{sect:oper}

The EP science observations have the following modes: monitoring survey, following-up observation, ToO observation, and instrument calibration. 
In addition, there will be instrument calibration observations for WXT and FXT.
The telescopes will always be pointed away from the Sun at a large avoidance angle. 

\begin{figure}[b]
\sidecaption[b]
\includegraphics[width=0.6\textwidth]{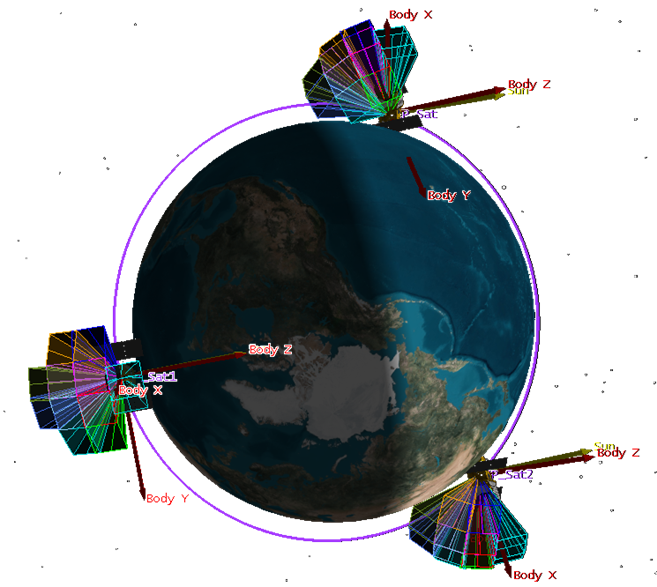}
\caption{Illustration of a pointing strategy for WXT in monitoring surveys. 
During one orbit 3 patches of the night sky will be observed with WXT. 
The entire night sky will almost be covered in three orbits (credit: IAMC, CAS).}
\label{fig:in-orbit}       
\end{figure}

\runinhead{Monitoring survey} 

This is the basic operation mode of EP, designed to monitor a large fraction of the sky. 
During one orbit, a few (currently set to three, as illustrated in Fig.~\ref{fig:in-orbit}) patches of the night sky will be observed with WXT as pointings, each with an exposure time ranging from  about 15 to 20 minutes.
While WXT taking data, FXT will normally be pointed to a pre-selected target of interest near its nominal pointing position, whenever possible, to maximise the science return. 
Only the night sky is monitored, which also allows prompt follow-up observations of the detected transients with ground-based optical/IR telescopes. 

\begin{figure}[b]
\sidecaption[b]
\includegraphics[width=1.\textwidth]{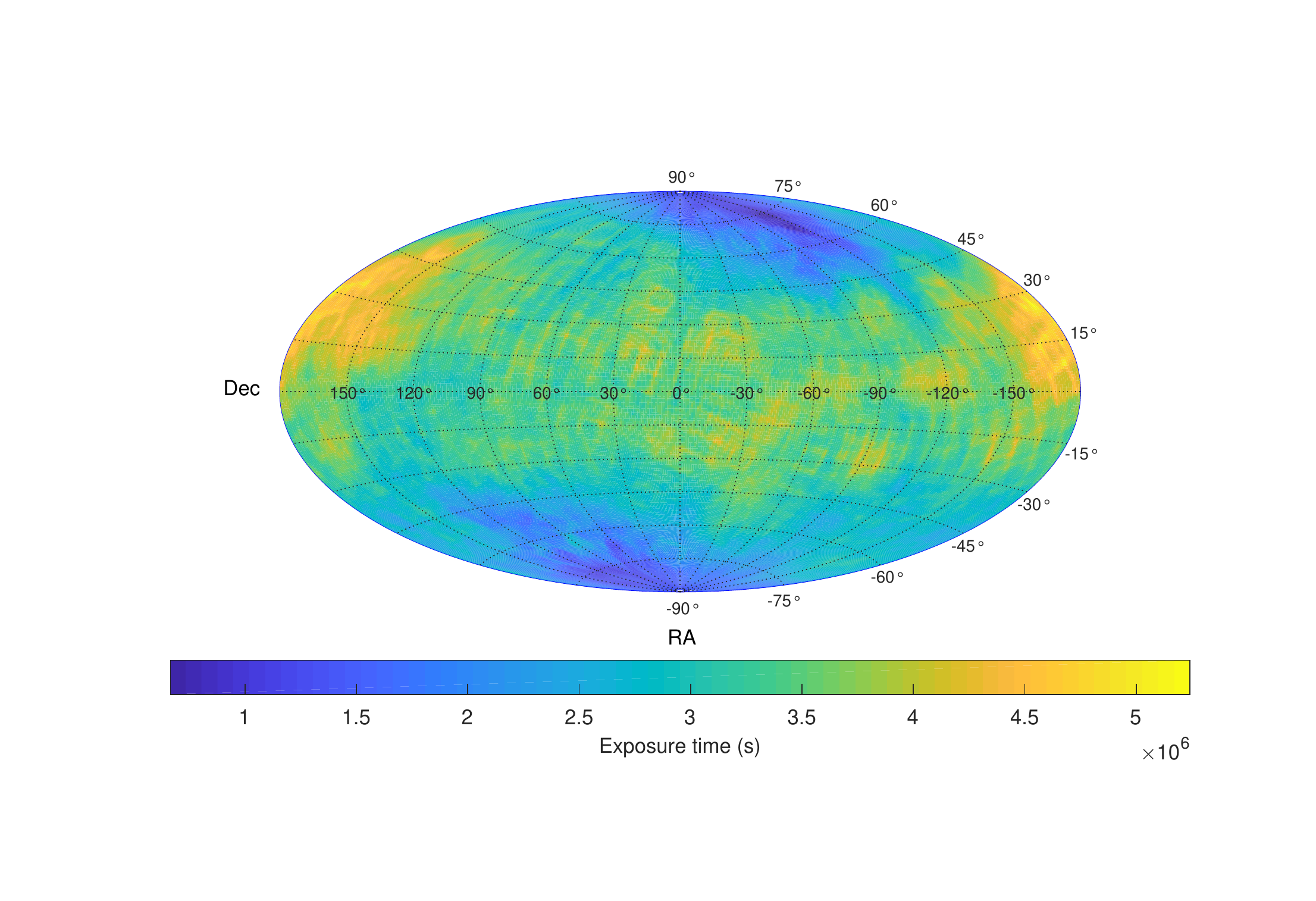}
\caption{Exposure map of WXT observations after one-year data taking for a chosen survey plan with three pointings per orbit in the night sky. Overheads are taken into account, including satellite slews and the SAA region, etc. (Pan et al. in prep.). }
\label{fig:expo}       
\end{figure}

\begin{figure}[t]
\sidecaption[t]
\includegraphics[width=1.\textwidth]{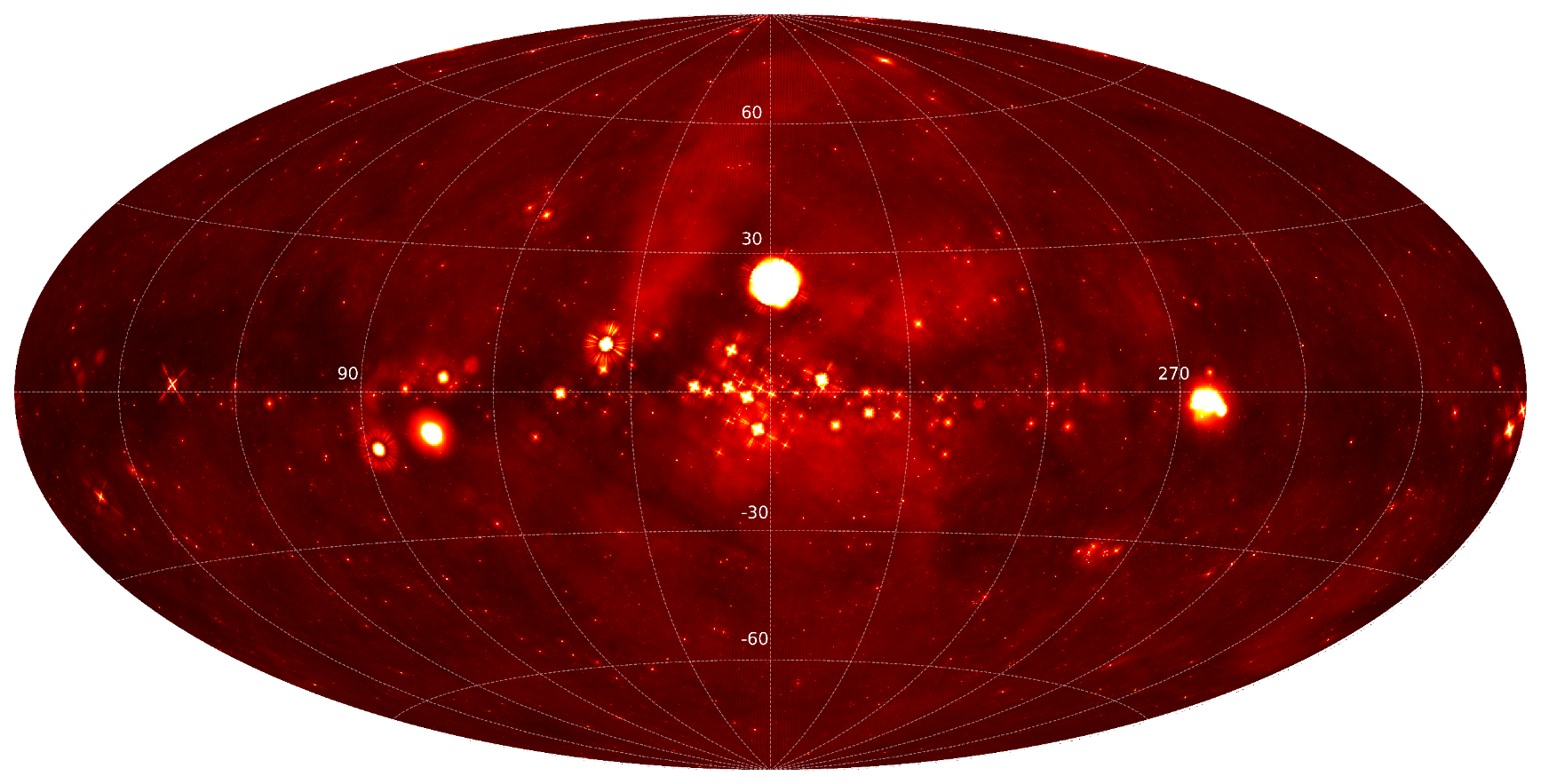}
\caption{Simulated all-sky X-ray map (in Galactic coordinates) from accumulated observations obtained with WXT in one-year data taking based on the above survey scheme (Pan et al. in prep.).}
\label{fig:epmap}       
\end{figure}

Over three successive orbits almost the entire night sky will be covered.
In this way, the sampling cadence for a given sky region ranges from 5 to 25 times per day.  
On average, the set of pointing directions is gradually shifted with time in accordance with the Sun's movement in the sky.
The whole sky can be covered in half-a-year operations.
EP will keep operating in this mode until a follow-up or ToO observation is triggered, after which the survey model is resumed. 
Fig.~\ref{fig:expo} shows an exposure map of WXT observations after one-year data taking for a survey plan with three pointings per orbit (assuming no follow-up or ToO observations).
Based on this survey scheme, the observational data (as photon events) taken with WXT in one-year data taking were simulated\footnote{The source catalogue and diffuse emission map from the RASS (Boller et al. 2016; Snowden et al. 1997), together with the MAXI source catalogue (Kawamuro  et al. 2018), were used as the inputs of the simulation.}, and an all-sky X-ray map produced from the simulated data is presented in Fig.~\ref{fig:epmap}.

\runinhead{Follow-up observation} 

Once a new transient is detected and meets a set of pre-defined criteria, 
a follow-up observation will be triggered. 
With automated and rapid slew of the spacecraft, FXT will be pointed to the position of the new source within 4 minutes normally, and start a pointed observation.
In general, follow-up observations last for a few orbits or more, with
the length of the observation being resettable via uplink command.  
Meanwhile, WXT continues to take data and monitor a new sky region around the target's position.

\runinhead{ToO observation} 
ToO observations will be carried out to observe interesting targets detected by EP, or by other multi-wavelength and multi-messenger facilities.
Both WXT and FXT can be utilised for ToO observations upon requests 
from the ground segment via the command data uplink route.
For normal ToOs the command will be uplinked within 24 hours, while for time-critical ToOs the uplink commands will be sent with a short latency (10 minutes or so as a goal). 
It is expected most ToO observations will utilise FXT, whereas WXT will continue to take data.

\subsection{Communications} 
\label{sect:com}

\begin{figure}[b]
\sidecaption[b]
\includegraphics[width=1.0\textwidth]{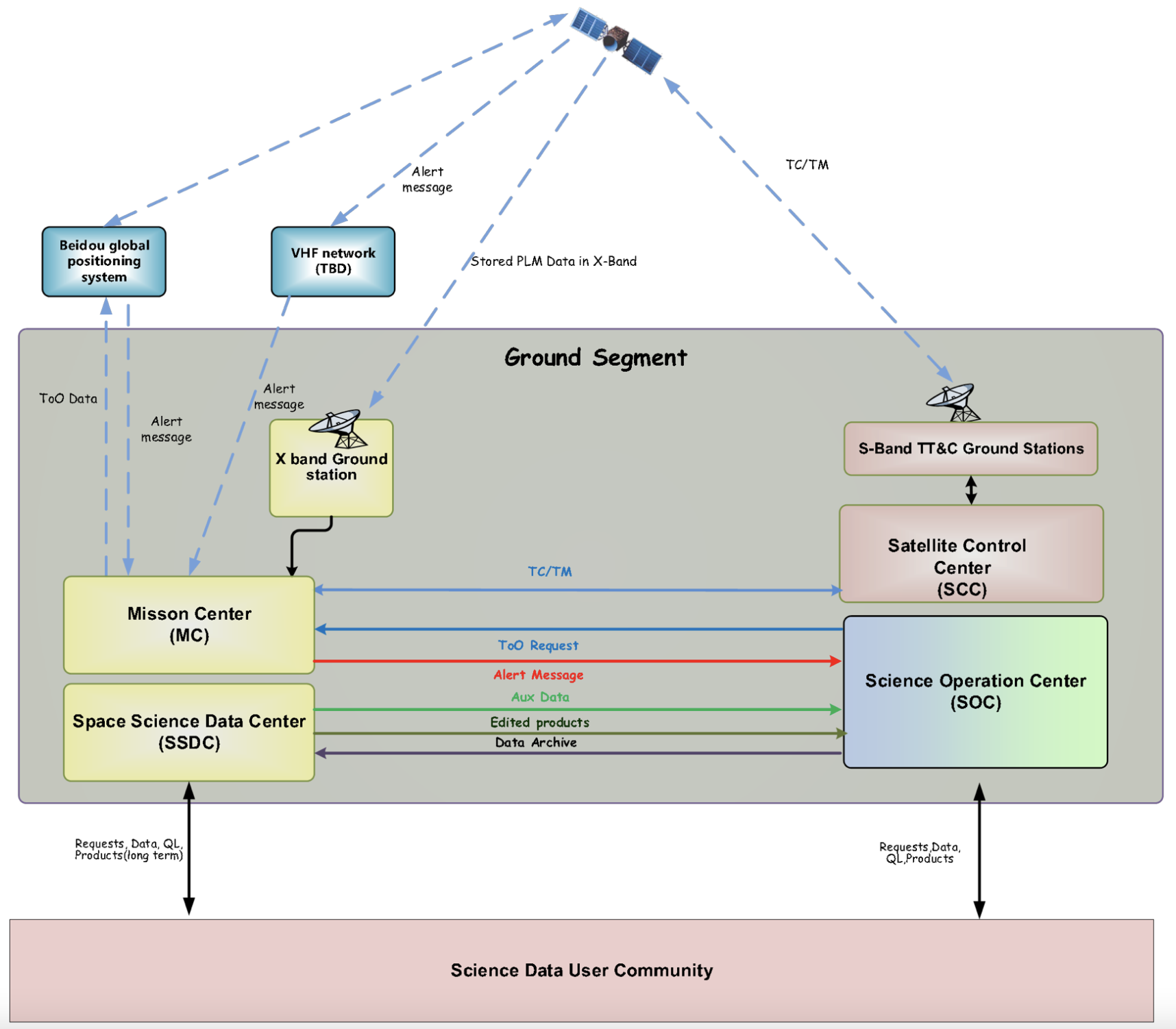}
\caption{Sketch of EP data and information flows and the ground segment (credit: NSSC, CAS).}
\label{fig:data}       
\end{figure}

A sketch of EP data and information flows and ground segment is shown in Fig.~\ref{fig:data}.

\runinhead{X-band data telemetry}
The scientific payloads and platform are expected to generate science and housekeeping data of about 134~Gbits per day. 
During satellite passages inside China, the Sanya station at Hainan will be used; whereas for other passages beyond the reach of Sanya the data downlink will be provided by other ground stations made available via the ESA contribution.
These ground stations are responsible for the X-band downlink, demodulating, descrambling, decoding, and data recording of EP data, as well as their transmission to the mission centre. 
The majority of the data generated by the satellite will be transmitted to the Mission Centre and Science Operation Centre within a few hours.  

\runinhead{Fast alert downlink and command uplink}

Upon detection of a transient source, its alert information will be downlinked to the ground segment with a latency about 10 minutes or so.
The alert data includes the coordinates, flux, spectral hardness ratio, and possibly a simple lightcurve of the source.
To do this, the \emph{Beidou} \footnote{The Chinese satellite navigation system} system will be utilised by taking advantage of its short text message capability and global coverage. 
To enhance the alert capability to transmit more of the quick-look data of transients, the VHF network system of the CNES (France), which is built for the Chinese-French SVOM mission, will be used via collaboration. 
The quick-look data will be helpful for the assessment and diagnostics of the transients.
For time critical ToO observations, the \emph{Beidou} system will be used to send uplink commands with short latencies, which will enable time-critical ToO observations such as a search for electromagnetic-wave sources of gravitational-wave events. 

\runinhead{TT\&C subsystem}

The TT\&C subsystem is responsible for undertaking the following tasks: orbit determination and orbit data uploading, remote monitoring of the status of the satellite and payloads, remote control of the operation modes of the satellite and payloads, commands uploading, time management of the satellite system, uploading of the satellite and payload parameters. 
For normal ToOs the commands will be uplinked via the regular S-band T/C route within 24 hours.

\subsection{Ground segment and science data}

The EP ground segment includes the Mission Centre (MC), Space Science Data Centre (SSDC), Science Operation Centre (SOC), Satellite Control Centre (SCC), X-band stations, S-band stations, and Data Communication Network (DCN). 
The MC, hosted at NSSC, is responsible for integrated operations and management, including mission planning and scheduling. 
The SOC, hosted mainly at NAOC and partly at IHEP, is responsible for the science operations of the instruments, including planning the science observing strategy, instrument status monitoring, data processing, handling ToO requests and user support. 
The SOC is designed to receive and respond to transient alerts and ToO requests, and to issue transient alerts to the worldwide community.
The SSDC at NSSC is responsible for science data pre-processing, data management, and permanent data archiving and formal public release.


The science data consist of X-ray events and calibration data generated by both WXT and FXT, as well as alert data produced from them.
The data of WXT and FXT observations will be processed through the data processing pipelines at the SOC located at NAOC and IHEP, respectively.
Corrected and cleaned events data (Level 2) and higher-level (Level 3) data products such as lists of detected sources  and their images, light curves, and spectra.
The transient alert message comprises the basic properties, such as the triggering time, significance of detection, source position, fluxes, spectral shape, timescale.
The alert messages will be made public immediately, after quick manual validation, to the global community to call for prompt multi-wavelength follow-up observations and identification.

The observational data, except for alert messages, will be studied internally by the EP science team within a proprietary period of generally one year, after which they will become publicly available. 
The proprietary period of data taken in ToO observations will be much shorter (less than six months).
There will be the Director Discretionary Time open to external users from the wider community, and data obtained via such a programme will be made publicly available immediately.
In addition, synergetic observations between EP and external observing facilities are anticipated, for which the resultant data will be jointly exploited via collaboration.

\begin{acknowledgement}
The authors would like to thank the colleagues of the EP team from the CAS, ESA and MPE who helped provide some of the contents and figures of the paper.
Special thanks are given to D. Zhao, Y. Liu, H. Pan, J. Zhang, Z. Cai, V. Burwitz, and P. Friedrich.
The work is supported by the Strategic Priority Research Program of the Chinese Academy of Sciences, Grant No. XDB23040100.
\end{acknowledgement}


\begin{thebibliography}{99.}%

\bibitem{abb17} Abbott et al. 2017, ApJL, 848, L13
\bibitem{ago03} Agostinelli, S., et al., 2003, NIMA 506, 250
\bibitem{arc21} Arcodia et al. 2021 Nature 592, 704
\bibitem{angel79} Angel J.R.P. 1979 ApJ, 233, 364
\bibitem{bol16} Boller Th., et al. 2016 A\&A, 558, A103
\bibitem{bui09} Buis E.J., Vacanti G. 2009 NIMA 599, 260
\bibitem{Chap91} Chapman H., et al., 1991 Review of scientific instruments 62(6), 1542
\bibitem{Chen20} Chen Y., Cui, W.W., Han, D.W., et al. 2020,  \emph{Proc. of SPIE.} 11444, 114445B
\bibitem{col15} Collier M.R., et al. 2015, Review of Scientific Instruments 86, 071301; doi: 10.1063/1.4927259 
\bibitem{cos97} Costa, E., et al. 1997, Nature, 387, 783 
\bibitem{fal78} Falk S.W. 1978 ApJL, 225, L133 
\bibitem{fra89} Fraser G.W. 1989, \textit{X-ray detectors in astronomy}  Cambridge Univ. Press
\bibitem{fra92}  Fraser G.W., et al., 1992 SPIE, 1546, 41
\bibitem{fra02} Fraser G.W., et al., 2002 SPIE 4497, 115 
\bibitem{fra10} Fraser G.W., et al., 2010 Planet. \& Space Sci. 58, 79
\bibitem{Fried2014} Friedrich P., et al. 2014 SPIE, 9144, 91444R
\bibitem{geh04} Gehrels N., et al. 2004 ApJ 611, 1005 
\bibitem{geh15} Gehrels N. \& Cannizzo J.K., 2015 JHEAp, 7, 2
\bibitem{Goet14} G\"otz D., et al., 2014 SPIE 9144, 23 
\bibitem{HP87} Holt S.S. \& Priedhorsky W.C. 1987 Space Sci. Rev., 45, 269
\bibitem{kaw18} Kawamuro T., et al. 2018 ApJS, 238, 32
\bibitem{kaa92} Kaaret P., et al., 1992 Appl. Opt. 31, 7339 
\bibitem{kb99} Komossa S. \& Bade N., 1999,  A\&A, 343, 775 
\bibitem{kom17} Komossa S., 2017, Astron. Nachr., 338, 256
\bibitem{lic21} Li C.K., et al. 2021, Nature Astro., 5,378
\bibitem{2009PASJ...61..999M} Matsuoka M. et al. 2009, PASJ, 61, 999
\bibitem{Meid2006} Meidinger N., et al. 2006 NIMA, 568, 141
\bibitem{met17} Metzger B.D. 2017 arXiv:1710.05931
\bibitem{mih11} Mihara T., et al. 2011, PASJ 63, S623 
\bibitem{obr21} O'Brien P., et al. 2021 Proceedings of the SPIE 2020, paper 11444-304 (arXiv:2102.08700) 
\bibitem{Pred2021} Predehl P., Andritschke R., Arefiev V., et al. 2021, A\&A, 647, A1
\bibitem{Prie96} Priedhorsky W.C., et al., 1996 MNRAS 279, 733
\bibitem{rees88} Rees M.J., 1988 Nature, 33, 523
\bibitem{saz21} Sazonov S., et al., 2021 MNRAS 508, 3820
\bibitem{sno97} Snowden S.L., et al., 1997 ApJ, 485, 125
\bibitem{sod08} Soderberg A.M., et al., 2008 Natur, 453, 469
\bibitem{sun19} Sun H., et al. 2019 ApJ, 886, 129
\bibitem{wang2020} Wang L., Qin L., Cheng J., et al. 2020, IEEE Transactions on Applied Superconductivity, PP(99):1-1
\bibitem{wil89} Wilkins S.W., et al., 1989 Review of scientific instruments 60(6), 1026
\bibitem{wil16} Willingale R., et al., 2016 SPIE 9905, 1
\bibitem{win03} Winkler C., et al. 2003 A\&A, 411, L1 
\bibitem{yuan15} Yuan W., et al., 2015 in PoS, \emph{Proceedings of "Swift: 10 Years of Discovery"}  arXiv:1506.07735
\bibitem{yuan16} Yuan W., et al., 2016 Space Science Review, 202, 235 
\bibitem{yuan18} Yuan W., et al., 2018 SPIE 10699, 1069925
\bibitem{zhang13} Zhang B. 2013 ApJL, 763, 22 
\bibitem{zhangj} Zhang, J., Qi L., Yang Y. et al., 2022, Astroparticle Physics. 137, 102668
\bibitem{zhao14} Zhao D.,  Zhang C., Yuan W., et al., 2014  SPIE, 9144, 91444E 
\bibitem{zhao2017} Zhao D., Zhang C., Yuan W., et al., 2017 Exp. Astron. 43, 267
\bibitem{zhao2018} Zhao D., Zhang C., Ling Z., et al., 2018 SPIE, 10699, doi: 10.1117/12.2311914


%
%
%
%
%

\end{thebibliography}
\end{document}